\documentclass[aps,reprint,article,superscriptaddress,twocolumn,showkeys]{revtex4-1}
\usepackage{blindtext}
\usepackage{amsmath}
\usepackage{amsfonts}
\usepackage{graphicx}
\usepackage{float}
\usepackage{booktabs}
\usepackage{gensymb}
\usepackage{hyperref}
\usepackage{braket}
\usepackage{soul}
\usepackage{svg}
\usepackage{amsthm}
\usepackage{amssymb}
\usepackage{array}
\newcolumntype{x}[1]{>{\centering\arraybackslash\hspace{0pt}}p{#1}}
\usepackage{soul} 

\usepackage{dsfont} 
\usepackage{MnSymbol}

\usepackage{color}

\definecolor{darkblue}{rgb}{0,0,0.5}

\hypersetup{
    bookmarksnumbered=true, 
    unicode=false, 
    pdfstartview={FitH}, 
    pdftitle={}, 
    pdfauthor={}, 
    pdfsubject={}, 
    pdfcreator={}, 
    pdfproducer={}, 
    pdfkeywords={}, 
    pdfnewwindow=true, 
    colorlinks=true, 
    linkcolor=darkblue, 
    citecolor=darkblue, 
    filecolor=darkblue, 
    urlcolor=darkblue 
}

\begin{document}

\title{Enhancing quantum models of stochastic processes with error mitigation}

\author{Matthew Ho}
\email{hosh0021@e.ntu.edu.sg}
\affiliation{Nanyang Quantum Hub, School of Physical and Mathematical Sciences, Nanyang Technological University, Singapore 637371, Singapore}

\author{Ryuji Takagi}
\email{ryuji.takagi@ntu.edu.sg}
\affiliation{Nanyang Quantum Hub, School of Physical and Mathematical Sciences, Nanyang Technological University, Singapore 637371, Singapore}

\author{Mile Gu}
\email{mgu@quantumcomplexity.org}
\affiliation{Nanyang Quantum Hub, School of Physical and Mathematical Sciences, Nanyang Technological University, Singapore 637371, Singapore}
\affiliation{Centre for Quantum Technologies. National University of Singapore, 3 Science Drive 2, Singapore 117543, Singapore}

\date{\today}

\begin{abstract}

Error mitigation has been one of the recently sought after methods to reduce the effects of noise when computation is performed on a noisy near-term quantum computer. Interest in simulating stochastic processes with quantum models gained popularity after being proven to require less memory than their classical counterparts. With previous work on quantum models focusing primarily on further compressing memory, this work branches out into the experimental scene; we aim to bridge the gap between theoretical quantum models and practical use with the inclusion of error mitigation methods. It is observed that error mitigation is successful in improving the resultant expectation values. While our results indicate that error mitigation work, we show that its methodology is ultimately constrained by hardware limitations in these quantum computers.

\end{abstract}


\maketitle

\section{Introduction}

Stochastic modelling is a key aspect of quantitative science, enabling us to simulate a process's potential behavior based on knowledge available in the present. In the spirit of Occam's Razor, there is often interest in minimal models -- models that replicate the conditional future of a relevant process while requiring minimal information about its past ~\cite{Crutchfield1989, Shalizi2001, Crutchfield2012a}. In this context, quantum computing can provide a marked advantage. Models that store the past in quantum memory can display a significant advantage -- replicating future statistics with less memory than all counterparts, and doing so in quantum superposition~\cite{ghafari2019interfering}. These advantages have led to new forms of quantum-enhanced dimensional reduction~\cite{Thompson2018,Elliott2020}, and the ability to create valuable superposition states that advantaged statistical analysis~\cite{Gu2012, Mahoney2016, Binder2018, Ho2020}. Indeed, the potential of both has been experimentally demonstrated in proof-of-principle spatially tailored optical quantum processors~\cite{Palsson2017, Jouneghani2017, Ghafari2018}.

With quantum computing gaining increasing attention alongside the availability of open source cloud quantum computing~\cite{Cross2017, cirq_developers_2021_4586899, Qutip1, Qutip2}, there are now exciting prospects to design such quantum models in much more general settings. However, great challenges remain. All results so far are predicated on near-perfect processors. Yet current Intermediate-Scale Quantum (NISQ) devices are noisy~\cite{Preskill2018quantumcomputing}, and it is expected that any result will be heavily influenced by noise. The traditional method to overcome such noise involves quantum error-correction codes~\cite{Steane1997, Gottesman1997, Steane2002, Gottesman2002}, but they typically sacrifice feasibility for accuracy by requiring large numbers of qubits. However, recent advances in error mitigation techniques provide an alternate method~\cite{Temme2017, Li2017efficient, McClean2017subspace, Bonet2018lowcost, Endo2018, McArdle2019error, koczor2021exponential, huggins2021virtual}, requiring far fewer qubits that make them attractive in the intermediate term.

Here, we introduce error mitigation methods to quantum modelling, making use of probabilistic error cancellation~\cite{Temme2017, Endo2018, Songeaaw5686}. To test its potential efficacy on present-day hardware, we make use of a real-world noise model extracted from \emph{ibmq\_toronto}, one of the IBM Quantum Falcon processors. We then simulate the execution of our quantum models when running on such hardware, with and without error mitigation, and benchmark their resulting accuracy using fidelity as a distance measure. Our numerical results indicate that error mitigation enables more accurate quantum models. We also investigate the use of limited shots in gate set tomography, the pre-experimental portion, and its effect on eventual simulation accuracy. Our work provides the ideal stepping stone for future quantum modeling on NISQ devices while reducing effects of noise.

\section{Framework}

\subsection{Stochastic processes and models}
The most intuitive form of a stochastic process takes the form of a time series. In the discrete scenario, imagine probing a system every $t$ time steps to observe the system's behaviour. The collection of outputs then forms the stochastic process. Each time step $t$ of the stochastic process is represented with a random variable $X_t$. $X$ can take values $x$ that reside in the set of alphabets $\mathcal{A}$.

Using subscripts to label the time indices, we then have stochastic process $\overleftrightarrow{X} \equiv X_{-\infty:\infty} \equiv  ..., X_{-1}, X_0, X_1, ...$ such that each instance of this process is $\overleftrightarrow{x} \equiv x_{-\infty:\infty} \equiv ..., x_{-1}, x_0, x_1, ...$. We denote each future by $\overrightarrow{x} = x_{0:\infty}$ and each past by $\overleftarrow{x} = x_{-\infty:0}$. Finite $L$-length sequences of a single instance of the time series is denoted $x_{0:L} \equiv x_{0}, x_{1}, ... , x_{0:L-1}$. Here, the left time index is inclusive and right time index is exclusive. A stochastic process is stationary if the statistical distribution of arbitrary-length sequences remains invariant, i.e., $P(x_{0:L}) = P(x_{\kappa:\kappa+L})$ for any $\kappa, L \in \mathbb{Z}^+$.

\subsection{Models}
\textbf{Classical models.}
Computational mechanics provides a statistical method for studying stationary stochastic processes \cite{Crutchfield1989, Shalizi2001, Crutchfield2012a} by producing the minimal model necessary to capture the dynamics of the stationary stochastic process. For an observer traversing the time series with full knowledge of the past $\overleftarrow{x}$, what information would the observer need track about $\overleftarrow{x}$ at each time-step, such that they can sequentially generate future predictions $x_1$, $x_2$, etc. that obey the desired conditional statistics $P( \overrightarrow{X} | \overleftarrow{x})$? 

One possibility is to retain all past outputs in memory. However, this involves a lot of waste, requiring unbounded memory to simulate a sequence of random bits. In the spirit of Occam's razor, computational mechanics provides the classically minimal alternative. It groups together pasts that have statistically identical futures together in equivalence classes, such that
\begin{equation}
\begin{aligned}
\label{eq.equivalencerelation}
	\overleftarrow{x} \sim \overleftarrow{x}' \iff P(\overrightarrow{X} = \overrightarrow{x} | \overleftarrow{x}) = P(\overrightarrow{X} = \overrightarrow{x} | \overleftarrow{x}').
\end{aligned}
\end{equation}
These equivalence classes are known as causal states, such that  pasts $\overleftarrow{x}$ and $\overleftarrow{x}'$ are equivalent if and only if $P(\overrightarrow{X} = \overrightarrow{x} | \overleftarrow{x}) = P(\overrightarrow{X} = \overrightarrow{x} | \overleftarrow{x}')$. Instead of tracking all past data, a model only needs to track which equivalence class it belongs to at each time-step,  drastically reducing memory requirements. Generating a sequence of random numbers with a model produced with computational mechanics, for example, would require no memory, as all pasts lead to statistically equivalent futures - alluding the observation that a machine that does this can blissfully throw unbiased coins at each time-step without tracking anything about what has transpired.

The set of causal states are denoted with $\{s_j\}$, where $j$ gives the index of the causal states (and not to be confused with the time index). Given that the past $\overleftarrow{x} \in s_j$ and at the next time step $\overleftarrow{x} x \in s_k$, we can label $s_k$ with a deterministic update function $\lambda(x,j)$. The deterministic function $\lambda(x,j)$ enables unifilarity; a model is unifilar if the next state is uniquely defined once $x$ is observed to be the output of $s_j$. The transition from $s_j$ to $s_{\lambda(x,j)}$ occurs probabilistically as output $x$ occurs with transition probability $P(x|j)$. The resultant model is thus represented by an edge-emitting hidden Markov model; nodes describe causal states and transitions given by directed edges. The edge-emitting hidden Markov model is also known as the process' $\varepsilon$-machine. 

The amount of memory required by the $\varepsilon$-machine is defined as the Shannon entropy of the stationary distribution of the causal states,
\begin{equation}
	C_\mu := H[P(s_j)] = -\sum_j P(s_j)\log_2 P(s_j).
\end{equation}
The quantity $C_\mu$ has been established as the (classical) statistical complexity of a process, an indicator of the memory requirement to simulate a stochastic process with the $\varepsilon$-machine. $C_\mu$ is considered a quantifier of structure in complexity science, measuring the fundamental resource costs needed for prediction ~\cite{Crutchfield1989, Crutchfield1997, Shalizi2008}. While $C_\mu$ clearly must be lower bounded by the past-future mutual information $E \equiv I(\overleftarrow{X},\overrightarrow{X})$, this bound is generally not strict~\cite{Crutchfield2009a}. There exist various methods of inferring the $\varepsilon$-machine from data~\cite{Crutchfield1989, Shalizi2004, Strelioff2014} and they have been widely applied to various fields to study neural spike trains~\cite{Haslinger2010}, dripping faucet experiments~\cite{Goncalves1998}, and stock markets~\cite{Park2007}.

\begin{figure}[t]
	\centering
	\includegraphics[width=1.0\linewidth]{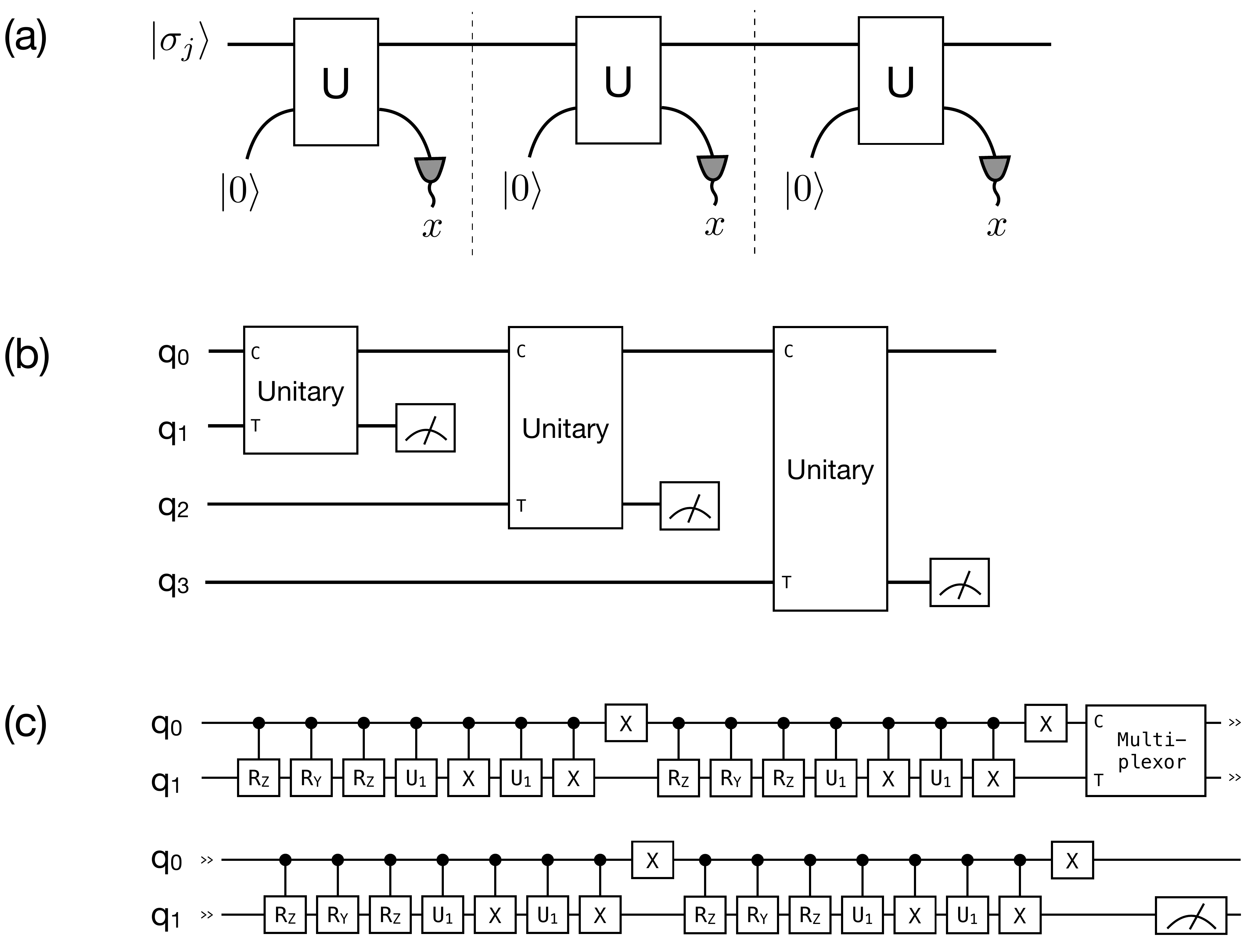}
	\caption{(a) The quantum model with the unitary acting on a quantum memory state $\ket{\sigma_j}$ for three time steps. (b) Original unitary operator tiled for 3 time steps. Here, $c$ labels the control qubit (memory register $\ket{\varsigma_i}$) while $T$ labels the target qubit (ancillary qubit $\ket{0}$). (c) Each unitary operator is decomposed using the cosine-sine decomposition into a series of controlled -$Z$ rotations, -$Y$ rotations, -NOTs, and -$U_1$ gates (single-qubit rotations about the $Z$-axis) interspersed with NOT gates. The gate labelled ``Multiplex" is a multiplexor with $T$ labelling the target qubit and $c$ as the control qubit.}
	\label{fig.decomposition}
\end{figure}

\textbf{Quantum models.}
Quantum mechanics enables quantum models of stochastic process that can reduce memory requirements below $C_\mu$~\cite{Gu2012, Tan2014, Mahoney2016, Binder2018, Liu2019, Ho2020}. They operate by mapping each causal state to a quantum memory state $s_j \mapsto \ket{\sigma_j}$ that satisfies the following relation under unitary evolution (see Fig.~\ref{fig.decomposition}(a)),
\begin{equation}
\label{eq.quantummemorystates}
	U \ket{\sigma_j} \ket{0} := \sum_{x\in\mathcal{A}} \sqrt{P(x|j)} e^{i\varphi_{xj}} \ket{\sigma_{\lambda(x,j)}} \ket{x},
\end{equation}
where $\varphi_{xj}$ is some tunable complex phase factor and $\ket{\sigma_j}$ forms the memory register~\cite{Binder2018, Liu2019}. The action of $U$ on the memory register with a blank ancillary qubit $\ket{0}$ changes the state of the memory register to $\ket{\sigma_{\lambda(x,j)}}$ while producing $\ket{x}$ with some probability $|\sqrt{P(x|j)}|^2$. Repeated applications of $U$ require either a new blank ancillary qubit or that the blank ancillary qubit be reset. This results in a sequence $\ket{x_0}\ket{x_1}\ket{x_2}...$ that gives the output string $x_0, x_1, x_2,... $ when measured in the appropriate basis. The output string will have the same statistical distribution as the input stochastic process.

The corresponding memory assigned to storing past information then corresponds to the von Neumann entropy computed as 
\begin{equation}
\label{eq.Cq}
	C_q := S[\rho] = -\text{tr}(\rho \log_2 \rho),
\end{equation}
where $\rho = \sum_j P(s_j) \ket{\sigma_j} \bra{\sigma_j}$ is the probabilistic mixture of quantum memory states weighted by their likelihood of occurrence. $C_q$ is referred to as the quantum statistical memory of the quantum model. In many processes, quantum models exist such that $C_q < C_\mu$, enabling enhanced quantum-memory advantage. Meanwhile continued execution of the quantum model without measuring outputs can lead to a quantum superposition of conditional futures~\cite{ghafari2019interfering}.

At present, there exists no systematic method to find the complex phases $\varphi_{xj}$ that minimise $C_q$. However, the special phase-less case where $\varphi_{xj} = 0$ can still lead to unbounded memory advantage~\cite{Elliott2019, Aghamohammadi2017, Garner2017b}. For the remainder of the paper, we consider such phase-less models, and refer to the resulting $C_q$ as the `quantum statistical memory' instead of `quantum statistical complexity'.

\textbf{Inferring Quantum Models.} Just as classical $\varepsilon$-machines can be inferred from the stochastic process, there exist protocols to infer the quantum models. One method would be to obtain the $\varepsilon$-machine using classical techniques, and then quantising the causal states using systematic techniques~\cite{Binder2018}. Meanwhile, the quantum inference protocol was recently developed to provide a direct approach inferring phase-less quantum models~\cite{Ho2020}. Here it was shown that if the chosen length of history to observe is at least the process' Markov order, $R$ \footnote{The Markov order $R$ is the smallest history length of the process to determine exact causal states, $H(S_R| X_{0:R}) = 0$.}, the quantum inference protocol is able to construct quantum memory states that have little variation from the exact quantum memory states. For situations where the exact $R$ is unknown, one may invoke the effective Markov order $R_\text{eff}$, which represents the minimal history length such that further pasts have little effect~\cite{Ho2020},
\begin{equation}
\begin{aligned}
\label{eq.effMarkovOrder}
	R_\text{eff} &= \text{min}\{r: \\
	&\text{max}_{x x'} \langle D(P(X_0|x X_{-r:0}), P(X_0|x' X_{-r:0})) \rangle < \xi \},
\end{aligned}
\end{equation}
where $D$ is some measure of distance between statistical distributions (such as trace distance) and $\xi$ is a tolerance value of choice, usually chosen as a function of the length of the stochastic process.

By using the stochastic process as the input, the quantum inference protocol constructs the following inferred phaseless quantum memory states from $L$-length of history $\ket{\varsigma_{x_{-L:0}}}$ which satisfies the following equation (see also Fig.~\ref{fig.decomposition}(a)),
\begin{equation}
\label{eq.inferredQMS}
	U \ket{\varsigma_{x_{-L:0}}} \ket{0} := \sum_{x_0\in\mathcal{A}} \sqrt{\tilde{P}(x_0|x_{-L:0})} \ket{\varsigma_{x_{-L+1:1}}} \ket{x_0}.
\end{equation}
A tilde is used to denote inferred quantities. The set of inferred quantum memory states has density matrix $\tilde{\rho}^{(L)} = \sum_{x_{-L:0}} \tilde{P}(x_{-L:0}) \ket{\varsigma_{x_{-L:0}}} \bra{\varsigma_{x_{-L:0}}}$. Likewise, the inferred quantum statistical memory is defined by the von Neumann entropy as
\begin{equation}
\label{eq.inferredCq}
	\tilde{C}_q := S[\tilde{\rho}^{(L)}] = -\text{tr}(\tilde{\rho} \log_2 \tilde{\rho}).
\end{equation}
Comparatively, it was shown that $\tilde{C}_q \to C_q$ if $L \geq R$ and when the length of the stochastic process, $N_\text{process}$, goes to infinity, i.e. $N_\text{process} \to \infty$. The quantum inference protocol was recently used to demonstrate the relative complexity of elementary cellular automata~\cite{Ho2021ECA}. 

The cardinality of $\mathcal{A}$ implies that the quantum inference protocol produces a maximum of $|\mathcal{A}|^L$ quantum memory states for $L$-length consecutive pasts. The state space increases exponentially with $L$, advocating a need to merge statistically similar quantum memory states to reduce the state space.

`Statistical similarity' between quantum states can be ascertained by computing the fidelity between two quantum states~\cite{Albuquerque2010}. The fidelity is essentially an equivalence relation analogous to Eq.~\eqref{eq.equivalencerelation}, taking into account the overlaps (inner products) between the quantum memory states, i.e. $|\braket{\psi|\phi}| = 1$ when $\ket{\psi} = \ket{\phi}$. Two quantum memory states $\ket{\varsigma_{x_{-L:0}}}$ and $\ket{\varsigma_{x'_{-L:0}}}$ are merged if the absolute value of their inner products are within some $\delta$-tolerance, i.e.
\begin{equation}
\label{eq.alternateequivalencerelation}
	\ket{\varsigma_{x_{-L:0}}} \sim \ket{\varsigma_{x'_{-L:0}}} \iff |\braket{\varsigma_{x_{-L:0}}|\varsigma_{x'_{-L:0}}}| \geq 1-\delta.
\end{equation}
$\delta$ should be a function of the length of stochastic process $N_\text{process}$ as a longer process begets better statistics to construct quantum memory states. We set $\delta = \frac{1}{2\sqrt{N_\text{process}}}$ and the derivation of $\delta$ is given in Supplementary Material~\ref{suppmat.fidelityQMS}. Merged quantum memory states are then relabelled as $\ket{\varsigma_i}$ before constructing the corresponding unitary operator. 

For any set of quantum memory states $\{\ket{\sigma_j}\}$, be it inferred or exact, there exists a systematic method to construct the appropriate unitary operator~\cite{Binder2018}. The full procedure is outlined in detail in Supplementary Material~\ref{suppmat.generatingunitaries}. In addition, we provide a systematic method to decompose the unitary operator into elementary quantum gates in Supplementary Material~\ref{suppmat.decomposingunitaries}.

\subsection{Quantum Error Mitigation}

\begin{figure}[t]
	\centering
	\includegraphics[width=1.0\linewidth]{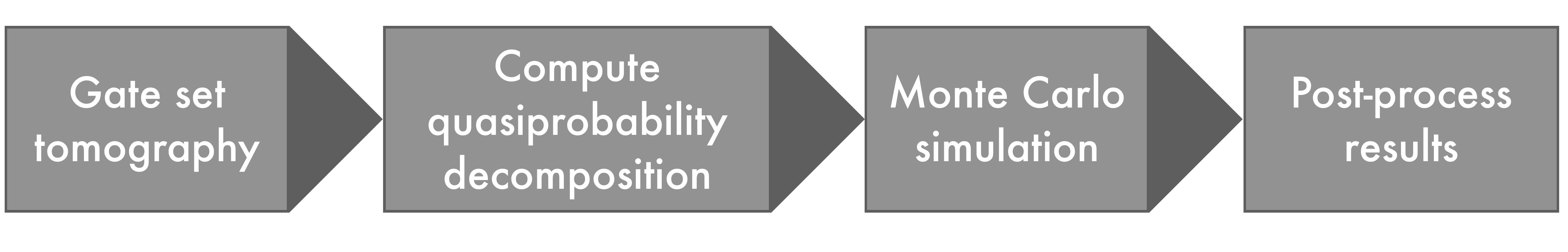}
	\caption{The error mitigation method consists of four steps as shown. (1) Gate set tomography is performed to obtain data that reconstructs the inverse noise after an operation $\bar{\mathcal{O}}$. (2) The gate set tomography data is post-processed to obtain the inverse noise, from which the inverse noise can be decomposed into basis operations with coefficients termed as quasiprobabilities. (3) By generating a circuit consisting of states, basis operations, and measurement basis with some probability according to the quasiprobabilities, Monte Carlo simulation can be carried out. (4) Lastly, the data from Monte Carlo simulation is post-processed.}
	\label{fig.QEM_workflow}
\end{figure}

Any quantum circuit on NISQ devices is bound to inherit errors in the calculations due to the presence of noise. Noise typically deviates the expectation values away from the ideal expectation value and error mitigation is used to reduce the amount of deviation. On this note, we run through the full methodology for error mitigation known as the probabilistic cancellation method, formalised in Ref.~\cite{Endo2018}. The workflow for the error mitigation method comprises of four key steps: (1) Gate set tomography, (2) Quasiprobability decomposition, (3) Monte Carlo simulation, and (4) Post-processing results. (See Fig.~\ref{fig.QEM_workflow}.)

We use the Pauli Transfer Matrix notation to describe the error mitigation technique we employ. Briefly, in the Pauli Transfer Matrix notation, a quantum state $\rho$ is transformed into a column vector 
\begin{equation}
    |\rho\rrangle \equiv [... \ \rho_\sigma \ ...]^T
\end{equation}
with each $\rho_\sigma$ is given by $\rho_\sigma = \text{tr}(\sigma \rho)$. Here, $\rho$ is the density matrix while $\sigma \in \set{I, \sigma_x, \sigma_y, \sigma_z}^{\otimes{n}}$ are the Pauli operators. $n$ denotes the number of qubits for the system.

Operators that act on $\rho$ transform $\rho$ into another state. These operators are quantum channels and are given by $\mathcal{O}(\rho) = \sum_k E_k \rho E_k^\dagger$, where $E_k$ are Kraus operators, $\sum_k E_k^\dagger E_k = I$. The resulting quantum channel $\mathcal{O}$ in Pauli Transfer Matrix notation is a real square matrix with elements
\begin{equation}
\label{eq.exactO}
	\mathcal{O}_{\sigma,\tau} = \frac{1}{2^n} \text{tr}[\sigma \mathcal{O} (\tau)]
\end{equation}
where $\sigma, \tau \in \set{I, \sigma_x, \sigma_y, \sigma_z}^{\otimes{n}}$ are Pauli operators. As such, $\rho' = \mathcal{O}(\rho)$ is akin to $|\rho' \rrangle = \mathcal{O} |\rho \rrangle$ in Pauli Transfer Matrix notation.

Measurements $M$ in the Pauli Transfer Matrix notation are row vectors
\begin{equation}
	\llangle M | \equiv [... \ M_\sigma \ ...]
\end{equation}
where $M_\sigma = \frac{1}{2^n}\text{tr}(\sigma M)$. $\sigma$ again is the set of Pauli operators. The expectation value is traditionally given as $\langle M \rangle = \text{tr}(M \rho)$. In the Pauli Transfer Matrix notation, $\langle M \rangle = \llangle M | \rho \rrangle$.

\textbf{Gate set tomography.}
Statistics of noise on the initial states, operations, and measurement bases are obtained via gate set tomography~\cite{Merkel2013self} that is performed with three steps. Consider the single-qubit case.

\begin{itemize}
	\item[(1)] Initialise a circuit with one of four linearly independent initial states $\bar{\rho}_k \in \set{\ket{0}\bra{0}, \ket{1}\bra{1}, \ket{+}\bra{+}, \ket{y+}\bra{y+}}$. $\ket{+} \equiv H\ket{0}$ is the eigenstate of the Pauli operator $\sigma_x$ and $\ket{y+} \equiv SH\ket{0}$ is the eigenstate of the Pauli operator $\sigma_y$, both with eigenvalue 1. Let $\bar{\rho}_{k=1} = \ket{0}\bra{0}$, $\bar{\rho}_{k=2} = \ket{1}\bra{1}$, $\bar{\rho}_{k=3} = \ket{+}\bra{+}$, and $\bar{\rho}_{k=4} = \ket{y+}\bra{y+}$.
	$H$ is the Hadamard gate while $S$ is a $\sqrt{Z}$ gate.
	\item[(2)] Insert the operator(s) $\bar{\mathcal{O}}^{(l)}$ that make up our circuit. (Each operator is labelled with superscript $(l)$ for more than one operator.)
	\item[(3)] Measure the expectation values for four linearly independent observables of the Pauli operators $\bar{M}_j \in \set{I, \sigma_x, \sigma_y, \sigma_z}$ for each quantum circuit. Specifically this means measuring the expectation values for $\langle \bar{M}_{j=1} \rangle = \langle I \rangle$, $ \langle \bar{M}_{j=2} \rangle = \langle \sigma_x \rangle$, $ \langle \bar{M}_{j=3} \rangle = \langle \sigma_y \rangle$, and $\langle \bar{M}_{j=4} \rangle = \langle \sigma_z \rangle$.
\end{itemize}

We use bars over variables to indicate that these elements of the quantum circuit are user-implemented. They would give ideal results should the quantum circuit be ideal but imperfect results when noise is involved.
A system of $n$ qubits demands $4^n$ initial states and $4^n$ measurement bases. When $n=2$ sixteen initial states $\bar{\rho}_k^{\otimes n} \in \set{\ket{0}\bra{0}, \ket{1}\bra{1}, \ket{+}\bra{+}, \ket{y+}\bra{y+}}^{\otimes n}$ and sixteen measurement bases $\bar{M}_j^{\otimes n} \in \set{I, \sigma_x, \sigma_y, \sigma_z}^{\otimes n}$ are required.

\textbf{Quasiprobability decomposition.}
To obtain the noise profiles for the initial states and measurement bases without operators, step (2) is omitted. This is used to compute a $2^n$ by $2^n$ Gram matrix $g$ with elements 
\begin{equation}
	g_{j,k} = \text{tr}(\bar{M}_j \bar{\rho}_k).
\end{equation}
The noise profile for each of the operators can be found by first computing the matrix $\tilde{\mathcal{O}}^{(l)}$ with elements
\begin{equation}
	\tilde{\mathcal{O}}^{(l)}_{j,k} = \text{tr}(\bar{M}_j \bar{\mathcal{O}}^{(l)} \bar{\rho}_k)
\end{equation}
before computing
\begin{equation}
\label{eq.Ohat}
	\hat{\mathcal{O}}^{(l)} = T g^{-1} \tilde{\mathcal{O}}^{(l)} T^{-1}
\end{equation}
using an invertible matrix $T = \begin{bmatrix} 1 & 1 & 1 & 1 \\ 0 & 0 & 1 & 0 \\ 0 & 0 & 0 & 1 \\ 1 & -1 & 0 & 0 \end{bmatrix}$ comprising of the Pauli operators basis sorted as $I, \sigma_x, \sigma_y, \sigma_z$, which the authors of Ref. \cite{Endo2018} recommend to be for error-free state preparation. Finally, the inverse noise $\left( \mathcal{N}^{(l)} \right)^{-1}$ of each operator can be calculated,
\begin{equation}
	\left( \mathcal{N}^{(l)} \right)^{-1} = \mathcal{O}^{(l),\text{exact}} \left( \hat{\mathcal{O}}^{(l)} \right)^{-1}.
\end{equation}
$\mathcal{O}^{(l),\text{exact}}$ is the exact Pauli Transfer Matrix representation of the operator(s) which can be found using Eq.~\eqref{eq.exactO}.
For small amounts of noise, $\left( \mathcal{N}^{(l)} \right)^{-1} \approx I$.

One can then find ways to implement the inverse noise after each $\bar{\mathcal{O}}^{(l)}$ is applied, effectively ``removing'' the noise which $\bar{\mathcal{O}}^{(l)}$ incurs. The inverse noise $\left( \mathcal{N}^{(l)} \right)^{-1}$ is decomposed and encoded into a set of basis operations $\hat{\mathcal{B}}_i$. These basis operations are applied directly after the original operator(s) $\hat{\mathcal{O}}^{(l)}$, essentially simulating the inverse noise of operator(s) $\hat{\mathcal{O}}^{(l)}$.

The set of basis operations that Ref. \cite{Endo2018} recommends requires post-selection. Post-selection limits the implementability of error mitigation as there needs to be a measurement mid-circuit with a conditional statement which verifies the circuit if some condition is met. If accepted, the circuit is left to continue running, else, the circuit is stopped and the next circuit initialised. 
However, measurements take much longer time than other quantum gates, making the error rates for measurements relatively high. 
In addition, the probabilistic nature of post-selection usually leads to a greater cost for probabilistic error cancellation~\cite{Takagi2020}.
Therefore, to circumvent this, we instead employ basis operations from Ref.~\cite{Takagi2020}, which only consists of deterministic operations and is applicable for any deterministic noise model. The $13$ basis operations for a single qubit system are listed in Table \ref{table.1} in Supplementary Material~\ref{suppmat.quantumerrormitigation}. We compute $\hat{\mathcal{B}}_i$ by replacing operations $\bar{\mathcal{O}}^{(l)}$ with $\bar{\mathcal{B}}_i$ in the three steps and perform the same computation in Eq.~\eqref{eq.Ohat}. Then, the inverse noise is encoded into a set of basis operations $\{ \hat{\mathcal{B}}_i \}$ with a quasiprobability decomposition:
\begin{equation}
	\left( \mathcal{N}^{(l)} \right)^{-1} = \sum_i q_{\mathcal{O}^{(l)},i} \hat{\mathcal{B}}_i
\end{equation}
Letting $|\hat{\rho}_k\rrangle = T_{\bullet,k}$ and $\llangle \hat{M}_j| = (g T^{-1})_{j,\bullet}$, where $M_{\bullet,k}$ is the $k^\text{th}$ column of matrix $M$ and $M_{j,\bullet}$ is the $j^\text{th}$ row of matrix $M$.
Quasiprobabilities of initial states and measurement bases are found as
\begin{equation}
	|\rho^{\text{exact}} \rrangle =  \sum_k q_{\rho,k} |\hat{\rho}_k\rrangle
\end{equation}
where $\rho^{\text{exact}} = \ket{0}\bra{0}$,
\begin{equation}
	\llangle M^{\text{exact}} | = \sum_j q_{M,j} \llangle \hat{M}_j |
\end{equation}	
where $M^{\text{exact}}$ is a $Z$-basis measurement.

Here, the quasiprobabilities $q_{\mathcal{O}^{(l)},i}$, $q_{\rho,k}$, and $q_{M,j}$ represent the coefficients for the basis operations, initial states, and measurement bases respectively.
We remark that one could choose a more general class of basis operations that include a continuous set of noisy implementable operations~\cite{Takagi2020,jiang2020physical,regula2021operational,Xiong2020sampling,piveteau2021quasiprobability}, which could reduce the sampling cost characterized in Eq.~\eqref{eq:cost}. We leave the detailed investigation on this extended basis operations as a future work.

\textbf{Monte Carlo simulation.}
For implementation using a Monte Carlo simulation, we compute the cumulative distribution functions of $q_{\mathcal{O}^{(l)},i}$, $q_{\rho,k}$, and $q_{M,j}$, also finding the following values in the process which are indicative of the sampling cost,
\begin{equation}
\begin{aligned}
    C_{\mathcal{O}^{(l)}} &= \sum_i | q_{\mathcal{O}^{(l)},i} | \\
    C_{\rho} &= \sum_k | q_{\rho,k} | \\
    C_M &= \sum_j | q_{M,j} |.
\end{aligned}
\label{eq:cost}
\end{equation}
Each Monte Carlo run comprises of a single shot: A circuit is initialised in state $\rho_k$ with some probability $P(q_{\rho,k}) = |q_{\rho,k}|/C_\rho$, operation $\bar{\mathcal{O}}^{(l)}$ applied, some basis operation $\hat{\mathcal{B}}_i$ chosen with probability $P(q_{\mathcal{B}_i}) = |q_{\mathcal{O}^{(l)},i}|/C_{\mathcal{O}^{(l)},i}$, then measured in basis $M_j$ with probability $P(q_M) = |q_{M,j}|/C_M$.  Each Monte Carlo run gives a measurement outcome $x \in \{0,1\}$. 

\textbf{Post-processing results.}
With $N_\text{MC}$ Monte Carlo runs, one can compute the estimated error mitigated expectation values as
\begin{equation}
\begin{aligned}
\label{eq.prob_QEM}
	&P_\text{QEM}(x=0) \\
	&= C \left( \frac{\text{No. of }x = 0 | \text{sgn} = +1}{N_\text{MC}} - \frac{\text{No. of }x = 0 | \text{sgn} = -1}{N_\text{MC}} \right) \\
	&P_\text{QEM}(x=1) \\
	&= C \left( \frac{\text{No. of }x = 1 | \text{sgn} = +1}{N_\text{MC}} - \frac{\text{No. of }x = 1 | \text{sgn} = -1}{N_\text{MC}} \right).
\end{aligned}
\end{equation}
Here, $\text{sgn}$ refers to the sign of the products of all coefficients used in each particular Monte Carlo run,
\begin{equation}
\begin{aligned}
	\text{sgn} &= \text{sign} \left( q_{\rho,k} [ \prod_l q_{\mathcal{O}^{(l)}} ] q_{M,j} \right) \\
	&= \text{sign} \left( q_{\rho,k} [ q_{\mathcal{O}^{(1)}} q_{\mathcal{O}^{(2)}} ... q_{\mathcal{O}^{(l)}} ] q_{M,j} \right).
\end{aligned}
\end{equation}
Also, $C$ is computed with
\begin{equation}
\begin{aligned}
\label{eq.C_coefficient}
	C &= C_{\rho} [ \prod_l C_{\mathcal{O}^{(l)}} ] C_M \\
	&= C_{\rho} [ C_{\mathcal{O}^{(1)}} C_{\mathcal{O}^{(2)}} ... C_{\mathcal{O}^{(l)}} ] C_M. 
\end{aligned}
\end{equation}

The Monte Carlo simulation is repeated many times until the standard deviation $\sigma_\text{MC} = \frac{C}{\sqrt{N_\text{MC}}}$ is within an acceptable value.

\textbf{Basis operations.}
Systems consisting of a single qubit require 13 basis operations~\cite{Takagi2020}.
These basis operations are CPTP maps and are listed in Table \ref{table.1}.
Systems with two qubits require $13^2 + 72 = 241$ basis operations.
While $13^2$ can be straightforwardly found by taking the tensor product between each $\bar{\mathcal{B}}_i$ in Table \ref{table.1}, the remaining $72$ operations are made up of CNOTs, controlled-phase, controlled-Hadamards, CNOTs with eigenstates of the Hadamard gate, SWAP, and iSWAP gates.
These 72 basis operations are listed in Table \ref{table.2}.

\section{Results}

\begin{figure}[t]
	\centering
	\includegraphics[width=0.8\linewidth]{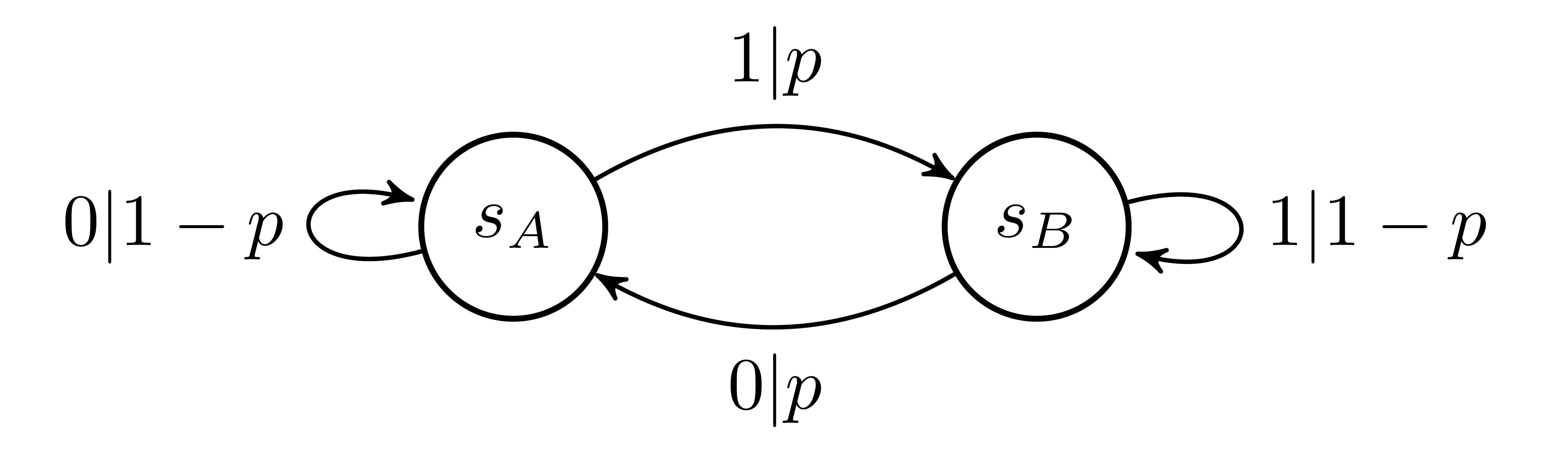}
	\caption{The perturbed coin generator is used to generate a bitstream of length $N_\text{process}$ with parameter $p$.}
	\label{fig.perturbedcoin}
\end{figure}

We are now equipped with the tools to execute an inferred quantum causal model's predictive capabilities on a quantum circuit, with and without error mitigation. With the stochastic process as the input, the quantum inference protocol is used to obtain quantum memory states directly. The unitary operator can be constructed and subsequently decomposed into elementary quantum gates. Lastly, error mitigation is employed to counter the effects of noise when generating future outputs.

Running the quantum circuit in Fig.~\ref{fig.decomposition}(c) is akin to applying the unitary operator once on the memory register $q_0$ and a blank ancillary qubit $q_1$ in Fig.~\ref{fig.decomposition}(b). This simulates using the memory register as a control qubit to perform controlled rotations on the target qubits $q_1$ before measuring $q_1$ in the appropriate basis. Multi-step outputs are attained by appending multiple blank ancillary qubits $q_1 = \ket{0}$, $q_2 = \ket{0}$, $q_3 = \ket{0}$ etc and allowing the memory register $q_0$ perform the necessary controlled rotations on qubit $q_1, q_2, q_3, ...$ and measuring them. In a perfect noiseless scenario, the measured outputs from qubits $q_1, q_2, q_3 ...$ yield $x_1, x_2, x_3, ...$ that will have a statistical distribution identical to the input stochastic process' distribution that is encoded within the memory register $\ket{\varsigma_j}$.

However, noise from the quantum computers will affect the rotations on the ancillary qubits, skewing the output distribution away from the exact distribution. As such, error mitigation is implemented to correct the rotations on the ancillary qubits, thus producing an output distribution that is closer to the exact distribution.

We demonstrate how quantum error mitigation can enhance quantum models when simulated on NISQ devices such as the IBM Quantum Experience. The quantum model is modelled after a stochastic process generated by the perturbed coin generator (Fig.~\ref{fig.perturbedcoin}). Due to its Markovian nature, the perturbed coin process provides a compact example for illustrating the error mitigation technique; constructing quantum memory states with the quantum inference protocol set at $L \geq 1$ would suffice, even when the length of stochastic process generated $N_\text{process}$ is in the order of $O(10^4)$, hence reducing statistical fluctuations in the probability amplitudes. The workflow would comprise of tuning the perturbed coin generator to a $p$-value ($0 < p < 1$, $p \neq 0.5$ \footnote{$p=0$ results in a period-2 process ...010101... while $p=1$ results in a period 1 process consisting of either all 1s or all 0s depending on which state is initialised. $p=0.5$ is in fact a random process.}) and generating a single shot of the stochastic process of arbitrary length $N_\text{process}$. ($N_\text{process}$ must not be too small or else the statistical distribution of the stochastic process itself would be inaccurate and affect the accuracy of the quantum memory states. Any $N_\text{process} \geq O(10^4)$ should suffice as $\delta \leq 1/\sqrt{10^4}$ in Eq.~\eqref{eq.alternateequivalencerelation}.) Once the phaseless quantum memory states are obtained, the unitary operator can be constructed and decomposed. The unitary operator would have dimensions $4$ by $4$ and one iteration of the cosine-sine decomposition would suffice.

As the perturbed coin process is effectively a two qubit system, we perform gate set tomography for the 16 initial states, 16 measurement bases, unitary operator, and the 241 basis operations to form the pre-experimental portion. The inverse noise of the unitary operator is then encoded into the 241 basis operations and Monte Carlo method is used for sampling. 

We make a brief note that we did not perform the error mitigation on the actual IBM Quantum Experience backend due to computational limits (time and resources), especially for the pre-experimental gate set tomography portion. Instead, the noise model from \emph{ibmq\_toronto} was extracted and used in the Qiskit simulator. The pros and cons of using such noise models will be discussed briefly in Sec.~\ref{sec.discussion}.

\begin{figure}[t]
	\centering
	\includegraphics[width=1\linewidth]{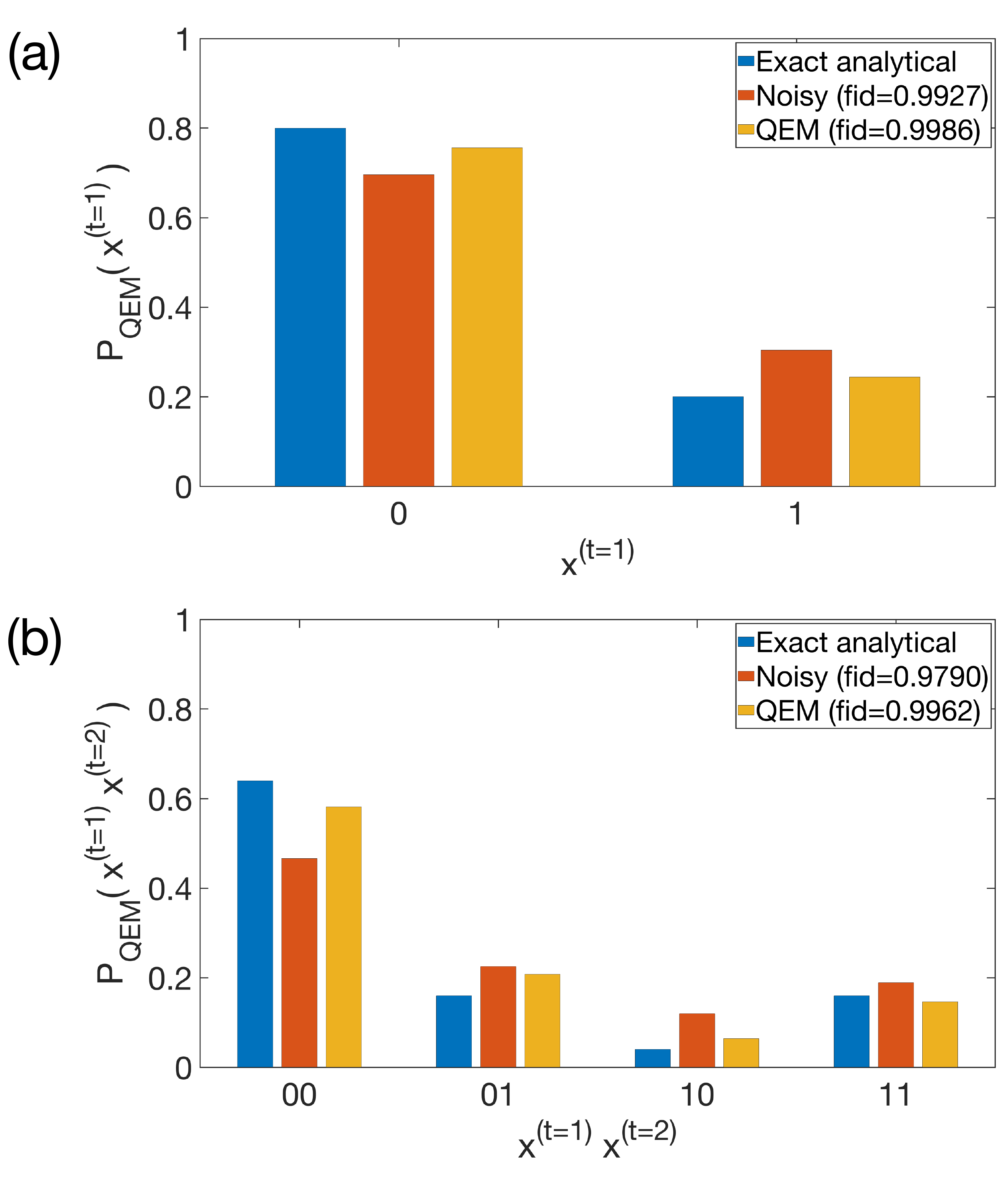}
	\caption{The multi-step joint statistical distributions for (a) 1 time step of unitary operation with outputs given by $x_\text{output} \equiv x^{(t=1)} \in \{0,1\}$ and (b) 2 time steps of unitary operator with outputs given by $x_\text{outputs} \equiv x^{(t=1)} x^{(t=2)} \in \{00, 01, 10, 11\}$, all of which are obtained with $N_\text{MC} = 10^7$ samples.}
	\label{fig.results_jointprob2}
\end{figure}

\textbf{Multi-step error mitigation.} Multi-step output results from the Monte Carlo simulation are post-processed to form a distribution in Fig.~\ref{fig.results_jointprob2}. A total of $N_\text{MC} = 10^7$ Monte Carlo trials, each trial resulting in one measurement outcome, are performed on the Qiskit simulator for the error mitigated quantum model. The same number of samples ($N_\text{noisy} = 10^7$) was obtained for a noisy non-error mitigated quantum model on the Qiskit simulator. Comparing the distributions of outcomes with the exact analytical solution, we observe that error mitigation improves the multi-step joint statistical distribution of the outcomes. One is free to select any choice of probability distance measure to quantify the closeness of two distributions to quantify the effectiveness of the error mitigation.

The Monte Carlo sampling with large $N_\text{MC}$ tends to produce a normal distribution for each outcome (i.e. $x^{(t=1)} \in \{0,1\}$, $x^{(t=2)} \in \{00,01,10,11\}$, and larger $t$) with standard deviation $\sigma_\text{MC} = \frac{C}{\sqrt{N_\text{MC}}}$. $C$ is the coefficient associated with the encoding of the inverse noise into basis operations and is proportional to the amount of noise in the quantum computer (i.e. large noise results in large $C$, see Eq.~\eqref{eq.C_coefficient}). For each outcome, applying the unitary operator $t$ times to obtain multi-step outputs incurs a compounded standard deviation of
\begin{equation}
\label{eq.multistep_error}
    \sigma^{(t)}_\text{MC} = \frac{C^t}{\sqrt{N_\text{MC}}}.
\end{equation} 

In our example, the unitary operator for the perturbed coin has $C \approx 20$. The corresponding standard deviation for a single time step output $x^{(t=1)}$ is $\sigma^{(t=1)}_\text{MC} \approx 20 / \sqrt{10^7} = 0.0063$, two time steps outputs $x^{(t=1)} x^{(t=2)}$ is $\sigma^{(t=2)}_\text{MC} \approx (20)^2 / \sqrt{10^7} = 0.1265$, and three time steps outputs $x^{(t=1)} x^{(t=2)} x^{(t=3)}$ is $\sigma^{(t=3)}_\text{MC} \approx (20)^3 / \sqrt{10^7} = 2.5298$. This straightforward calculation implies that 3 time steps of unitary for the perturbed coin scenario is currently computationally impractical, largely due to how big $C$ is; one would be required to have $N_\text{MC} \geq 6 \times 10^9$ Monte Carlo runs to have an error of roughly $10\%$.

The factor $C$ is highly dependent on the amount of noise in the quantum computer and is not within any immediate control. A large $C$ such as $C \approx 20$ in our example affects $\sigma_\text{MC}^{(t)}$. As quantum hardware improves over time, $C$ is expected to decrease, making error mitigation for larger multi-steps more feasible.

In fact, the factor $C$ is also affected by the number of shots used in the pre-experimental portion in gate set tomography, which we denote with $N_\text{GST}$. A small $N_\text{GST}$ causes $\left( \mathcal{N}^{(l)} \right)^{-1}$ to deviate from the ideal operation which will affect the quasiprobability decomposition into basis operations. The basis operations, which are part of gate set tomography, will have errors induced as well. These errors will compound and affect $C$. Ideally, one would choose a large $N_\text{GST}$ to ensure that the errors are kept to a minimum. However, due to quantum devices such as the IBM Quantum Experience having a limit to the number of shots, the accuracy of $\left( \mathcal{N}^{(l)} \right)^{-1}$ and $C$ will be affected. For our simulation, we follow the limits on the actual backend and keep $N_\text{GST} = 8192$ (the highest number of shots) and derive an approximate scaling for errors in $\left( \mathcal{N}^{(l)} \right)^{-1}$ and $C$ with respect to $N_\text{GST}$ in Supplementary Material~\ref{suppmat.quantumerrormitigation}.

\begin{figure*}[t]
	\centering
	\includegraphics[width=1\linewidth]{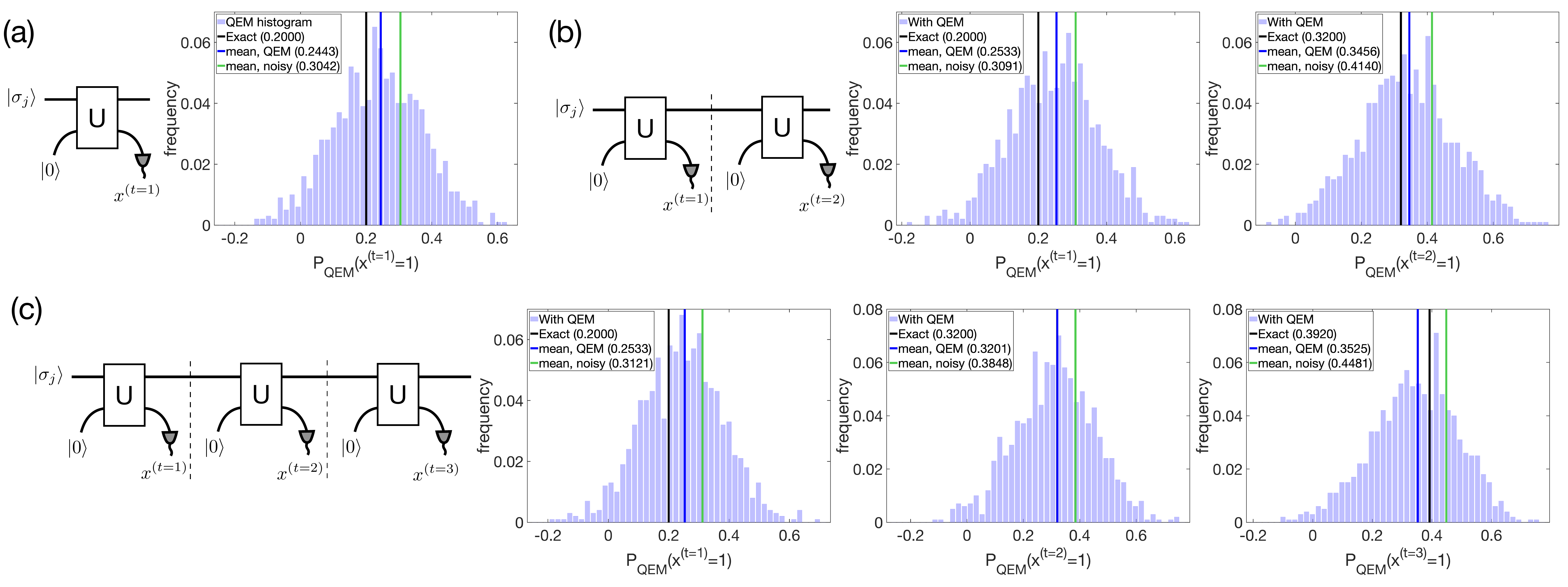}
	\caption{Distribution for the expectation values for $P_{\text{QEM}}(x^{(t)}=1)$ i.e. for outcome $x = 1$ for (a) t=1 time step of unitary operation, (b) t=2 time steps of unitary operations, and (c) t=3 time steps of unitary operations. All figures had error mitigation carried out (blue histogram) and the results for the output $1$ are compared against the ideal quantum simulator without noise (red) and quantum simulator with noise (green). The histograms are constructed with $10^3$ chunks, each chunk having $N_\text{MC,chunk} = 10^4$ samples.}
	\label{fig.results_indivoutputs}
\end{figure*}
Error mitigation for individual outputs can also be computed from the Monte Carlo samples. Since individual outputs are due to one instance of unitary operation, the corresponding standard deviation is independent of $t$,
\begin{equation}
\label{eq.singlestep_error}
    \sigma_\text{MC} = \frac{C}{\sqrt{N_\text{MC}}}.
\end{equation} 

\textbf{Individual-steps error mitigation}. Theoretically, depending on the size of $C$, a smaller sample size can be used to obtain decent statistics on individual outputs. For $C \approx 20$, $N_\text{MC} \geq 10^6$ is required to have $\sigma_\text{MC} \leq 0.02$, or roughly 2\% error. The disadvantage of individual steps is that it is difficult to reconstruct the multi-step joint probability distributions. This is because any correlation between individual outputs is abandoned when the results of individual outputs are processed. The multi-step joint probability distributions cannot be recovered unless the input stochastic process is a fully random process -- only then will the following equation hold,
\begin{equation}
    P(x_{0:K}) = P(x_0) P(x_1) P(x_2) ... P(x_{K-1}),
\end{equation}
where $x_{0:K}$ is the future outputs from $K$ unitary operations.

Nonetheless, with a total of $N_\text{MC} = 10^7$ in our dataset, we split these into a total of $10^3$ chunks with each chunk having $N_\text{MC,chunk} = 10^4$ samples. Splitting the entire dataset simulates running the Monte Carlo simulation in chunks and analysing each chunk individually. Each chunk then has its individual expected $P_\text{QEM}(x=1)$ computed before the collection of expected $P_\text{QEM}(x=1)$'s is plotted in a histogram to show the frequency of occurrence. With $N_\text{MC,chunk} = 10^4$, we expect the standard deviation of the histogram to be $\sigma_\text{MC,chunk} \approx \frac{20}{\sqrt{10^4}} = 0.2$.

In Fig.~\ref{fig.results_indivoutputs}, we illustrate how $P_\text{QEM}(x^{(t)}=1)$ is distributed with the mean of the distribution up to $t=3$. The mean is observed to be closer to the exact analytical value than the mean of $P_\text{noisy}(x^{(t)}=1)$ which indicates the success of the error mitigation method. However, because $N_\text{MC,chunk} = 10^4$, the distribution with $\sigma_\text{MC,chunk} \approx 0.2$ is reflected in the amount of spread.

\textbf{Memory costs.} The success of quantum error mitigation is illustrated in Figs. \ref{fig.results_jointprob2} and \ref{fig.results_indivoutputs}, but at what cost can they be counted as successful? Notably, entropic measures of memory take the greatest operational meaning in the context of simultaneous simulation. They represent the average memory needed to simulate each instance of a process when simulating a large number of such processes at the same time. In this context, we define the \emph{cumulative memory cost} as the total memory that an arbitrary model needs to build a distribution of the futures using $N_\text{model}$ samples where $N_\text{model} \gg 1$.

The $\varepsilon$-machine for the perturbed coin process has $C_\mu = 1$ at $0 \leq p \leq 1$ and $C_\mu = 0$ at $p=0.5$. The quantum causal models on the other hand have quantum statistical memory $C_q$ that is continuous in the range $0 \leq p \leq 1$, given by the following equation~\cite{Gu2012},
\begin{equation}
\begin{aligned}
	C_q = &- \left( \frac{1}{2} + \sqrt{p(1-p)} \right) \log_2 \left( \frac{1}{2} + \sqrt{p(1-p)} \right) \\
	&- \left( \frac{1}{2} - \sqrt{p(1-p)} \right) \log_2 \left( \frac{1}{2} - \sqrt{p(1-p)} \right).
\end{aligned}
\end{equation}
For instance, the $\varepsilon$-machine for the perturbed coin process requires $C_\mu = 1$ bit of memory (at $p=0.2$) while the unitary quantum models require $C_q = 0.4690$ bits of memory (at $p=0.2$). 

In a perfect noiseless scenario, the cumulative memory cost of quantum models would be less than that of the $\varepsilon$-machine as well. If we set the standard deviation of the output statistical distribution to be $\sigma = 0.02$, the $\varepsilon$-machine requires $N_\text{$\varepsilon$-machine} = 2500$ samples. The associated cumulative memory cost would be $2500 \times 1 = 2500$ bits. Similarly, (noiseless) quantum models require $N_\text{q-models} = 2500$ samples and its cumulative memory cost would be $2500 \times 0.4690 = 1172.5$ bits, showing a clear advantage in cumulative memory cost.

Simulating quantum models on NISQ devices such as the IBM Quantum Experience requires error mitigation due to noise from the quantum computers. Along this vein, for error mitigated quantum models to have a standard deviation of $\sigma_\text{MC} = 0.02$ as well, the number of samples would be $N_\text{MC} = \left( \frac{C}{\sigma_\text{MC}} \right)^2 = \left( \frac{20}{0.02} \right)^2 = 10^6$ samples. The corresponding cumulative memory cost for error mitigated quantum models would be $10^6 \times 0.4690 = 469000$ bits, far exceeding that of classical models, putting error mitigated quantum models at a clear disadvantage over $\varepsilon$-machines.

\begin{figure}[t]
	\centering
	\includegraphics[width=1.0\linewidth]{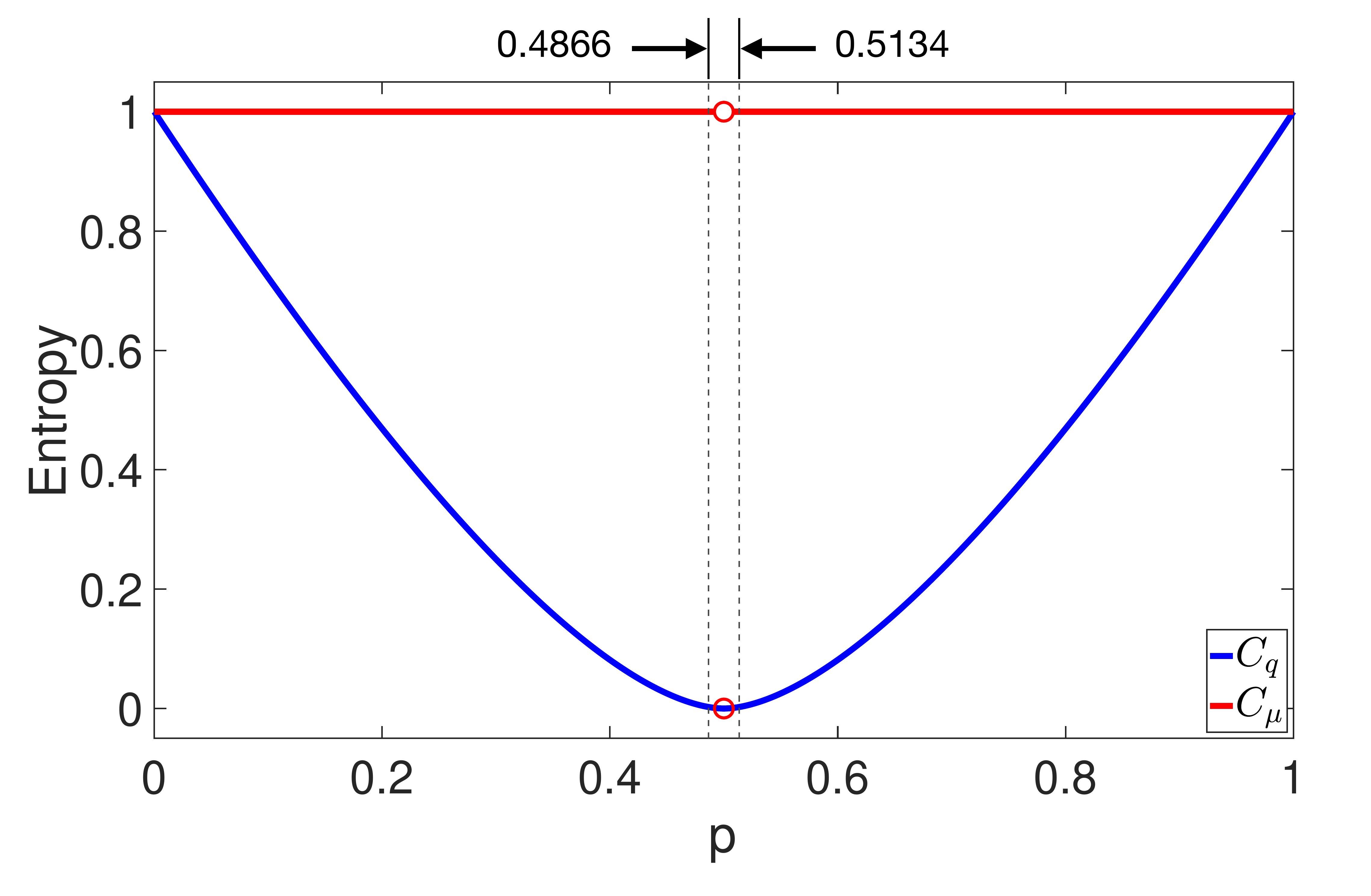}
	\caption{The perturbed coin has the following $C_\mu$ and $C_q$ for $0 \leq p \leq 1$. $C_\mu$ is discontinuous only at $p=0.5$ as the $\varepsilon$-machine at $p=0.5$ is essentially a random process that requires only a single causal state. The region where $0.4688 \leq p \leq 0.5134$ ($p \neq 0.5$) denotes the region where the cumulative memory cost for the quantum models require outdoes the memory cost for the $\varepsilon$-machine when error mitigation methods are taken into account.}
	\label{fig.CqvsCmu}
\end{figure}

The $p$-value indirectly controls the degree of non-orthogonality between quantum memory states which corresponds to a range of values of $C_q$. Given the current climate of noisy quantum computing where noise is highly dependent on hardware, we ask, how much non-orthogonality is required between quantum memory states before the cumulative memory cost of error mitigated quantum models matches and is lower than that of the $\varepsilon$-machine? In other words, what $p$-value(s) will give a cumulative memory advantage even with error mitigation? The solution is traced back to the quantum statistical memory of quantum models. We solve the following inequality to find the values where $C_q$ is able to reduce the cumulative memory cost by substituting $N_\text{MC} = 10^6$ and $N_\text{$\varepsilon$-machine} = 2500$,
\begin{equation}
\begin{aligned}
	N_\text{MC} \times C_q &\leq N_\text{$\varepsilon$-machine} \times C_\mu \\
	C_q &\leq 0.0025 \text{ bits}.
\end{aligned}
\end{equation}
The corresponding $p$-values for which this $C_q \leq 0.0025$ is $0.4866 \leq p \leq 0.5134$, $p \neq 0.5$. We illustrate this region between two vertical lines in a plot of $C_\mu$ and $C_q$ against $p$~\cite{Gu2012} in Fig.~\ref{fig.CqvsCmu}. As quantum hardware improves, reduction in noise will reduce the value of $C$ and therefore increase the range of $p$ values where we anticipate a quantum memory advantage.

\textbf{On feasibility of error mitigation.} The baseline sampling cost for error mitigation given by Eq.~\eqref{eq.singlestep_error} increases with $C$ and decreases with $N_\text{MC}$. $C$ is a hardware-dependent variable as it depends on how noisy a quantum computer is while $N_\text{MC}$ is time and resource dependent -- how many Monte Carlo samples are needed given the availability of running experiments on quantum computers.

We showed that with $C \approx 20$ and $N_\text{MC} = 10^7$, the standard deviation for future multi-step joint statistical distributions and fidelity quickly diverges, rendering three or more time steps of unitary operation unusable. The standard deviation with $N_\text{MC} = 10^6$ will suffice in obtaining a decent statistical distribution for individual outputs. This comes with a trade-off -- reconstructing the multi-step joint statistical distribution from individual distributions is not possible as correlations between individual outputs have been ignored in the post-processing phase of error mitigation. As such, there is a fine balance between sampling cost and distribution, only to be outweighed with any ignorance of correlations between the outputs.

\section{Discussion}
\label{sec.discussion}

Noise is a major impediment for realising quantum advantage in present-day quantum computers. In the context of quantum modelling, they distort future predictions and result in non-ideal predictions. Here, we illustrated how such distortions can be reduced through quantum error mitigation though at the cost of higher cumulative memory cost. We illustrated the efficacy of this technique via the case study on the perturbed coin process using an actual real-life noise model extracted from one of the IBM Quantum Falcon processors, \emph{ibmq\_toronto}. Error mitigation was then carried out on the unitary operator from the elementary quantum gates to reveal a statistical distribution that has better fidelity. Meanwhile, there remains a parameter regime where the quantum model's memory overhead is still sufficiently low to allow an overall quantum memory advantage even in present-day NISQ devices. We expect this parameter regime to widen as quantum hardware improves.

Simulators with noise models extracted from the real backend tend to overlook the non-Markovian nature of noise that are found in real quantum computers~\cite{morris2019nonmarkovian, chen2020nonmarkovian}. However, a noise model extracted from the real backend provides a more realistic representation of noise as compared to custom noise such as depolarising noise, dephasing noise, and amplitude damping. This paper hence illustrates the potential of enhancing quantum models of stochastic processes with error mitigation regardless of noise, providing a stepping stone for a potential solution to dealing with noise on NISQ devices.

In addition to the memory consideration, a second key advantage of quantum models is their capacity to generate conditional future distributions in quantum superposition~\cite{ghafari2019interfering}. This superposition state then forms a key resource for quantum amplitude estimation protocols that are key for quantum-enhanced analysis of stochastic data~\cite{brassard2002quantum,blank2020quantum}. In this instance, the extra memory overhead due to error mitigation is not necessarily a significant issue, as it simply reflects the necessity to run simulation a certain extra number of times. Thus, one exciting future direction is to ascertain the regimes where there are advantages of generating superposition of possible futures.

\section{Acknowledgements}

We acknowledge the use of IBM Quantum services for this work. This work is supported by the the Singapore Ministry of Education Tier 1 grant 2019-T1-002-015 (RG190/17), the National Research Foundation Fellowship NRF-NRFF2016-02, the Quantum Engineering Program QEP-SP3, and the FQXi-RFP-1809 from the Foundational Questions Institute and Fetzer Franklin Fund, a donor advised fund of Silicon Valley Community Foundation. The views expressed are those of the authors and do not reflect the official policy or position of IBM and the IBM Quantum team or the National Research Foundation of Singapore.

\bibliographystyle{apsrmp4-2}
\bibliography{QEM}
 

\clearpage

\pagebreak
\widetext

\begin{center}
\textbf{\large Supplementary Material: Enhancing quantum models of stochastic processes with error mitigation}
\end{center}

\begin{center}
Matthew Ho$,^{1}$ Ryuji Takagi$,^{1}$ and Mile Gu${}^{1,\;2}$

\emph{\small ${}^{\mathit{1}}$Nanyang Quantum Hub, School of Physical and Mathematical Sciences, Nanyang Technological University, Singapore 637371, Singapore\\
${}^{\mathit{2}}$Centre for Quantum Technologies, National University of Singapore, 3 Science Drive 2, Singapore 117543, Singapore}
\end{center}

\setcounter{equation}{0}
\setcounter{figure}{0}
\setcounter{table}{0}
\setcounter{page}{1}
\setcounter{section}{0}
\renewcommand{\theequation}{S\arabic{equation}}
\renewcommand{\thefigure}{S\arabic{figure}}
\renewcommand{\thepage}{S\arabic{page}}
\renewcommand\thesection{\Alph{section}}

\section{Causal models}
\label{sec.causalmodels}

Occam's Razor is the driving principle behind the construction of causal models. In general modelling terms, the simplest model is usually the right one. In causal modelling, the adjective `simple' points at the model not needing unnecessary information of the past to predict the future.

Take a random process for example. One may feel compelled to remember how all past trajectories lead to all futures to learn of the process' dynamics in order to do prediction. However, causal modelling indicates that since all pasts lead to the same statistical futures, there is redundancy in remembering all possible pasts, and may be grouped together in the same equivalence class.

Causal models abide by the equivalence relation which states that all pasts with the same statistical futures should be merged into an equivalence class,
\begin{equation}
\begin{aligned}
\label{eq.equivalencerelation_suppmat}
	\overleftarrow{x} \sim \overleftarrow{x}' \iff P(\overrightarrow{X} = \overrightarrow{x} | \overleftarrow{x}) = P(\overrightarrow{X} = \overrightarrow{x} | \overleftarrow{x}').
\end{aligned}
\end{equation}

\begin{figure}[!h]
	\centering
	\includegraphics[width=0.35\linewidth]{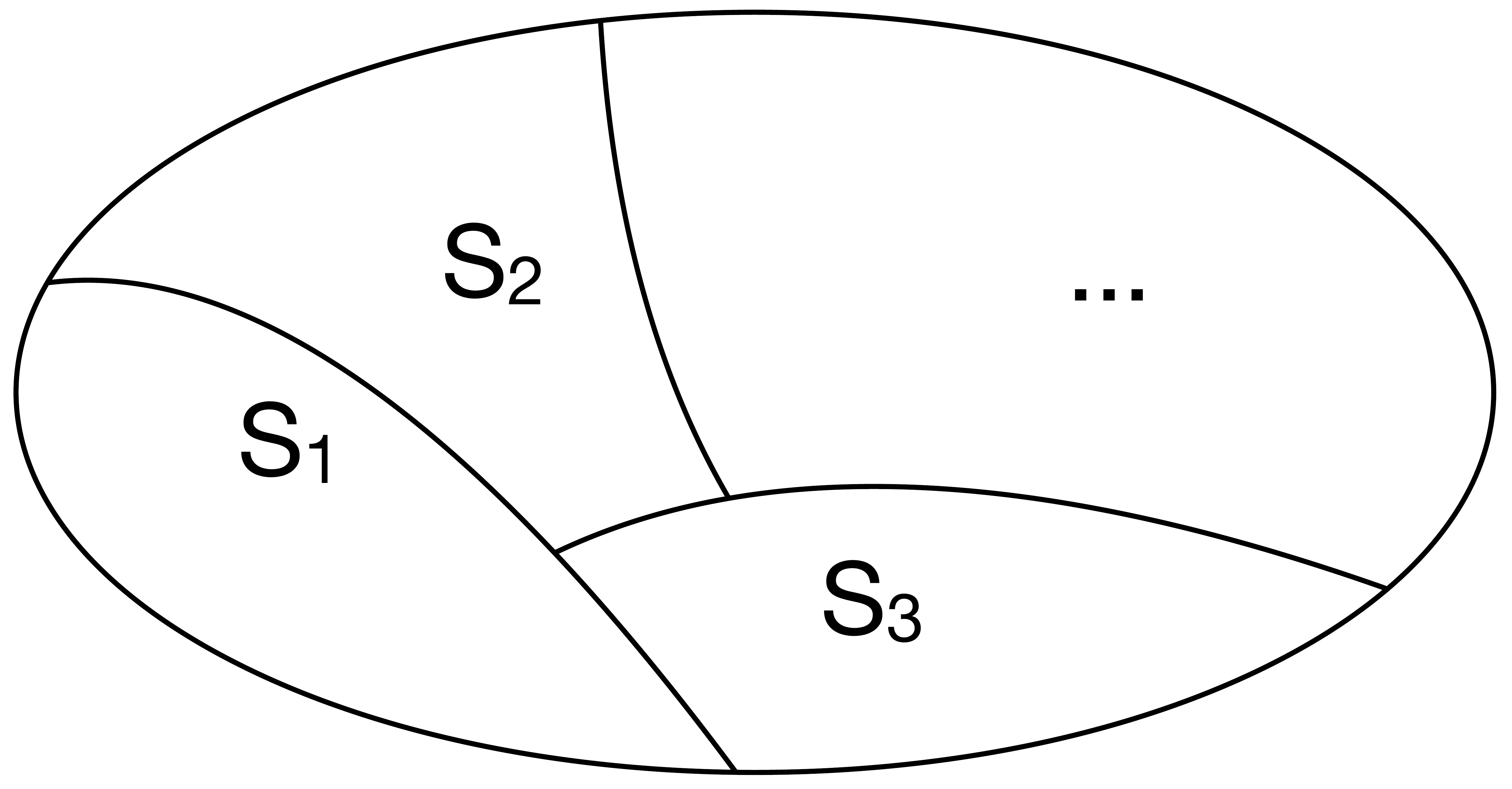}
	\caption{Equivalent pasts as determined by the equivalence relation are grouped together in an equivalence class. The set of equivalence classes forms the state space of all pasts.}
	\label{fig.equivalenceclasses}
\end{figure}

Invoking Eq.~\eqref{eq.equivalencerelation_suppmat} effectively sieves out all equivalent pasts, grouping them together. The state space of all pasts is reduced to groups of equivalent pasts illustrated in Fig.~\ref{fig.equivalenceclasses}.

\begin{figure}[!h]
	\centering
	\includegraphics[width=0.4\linewidth]{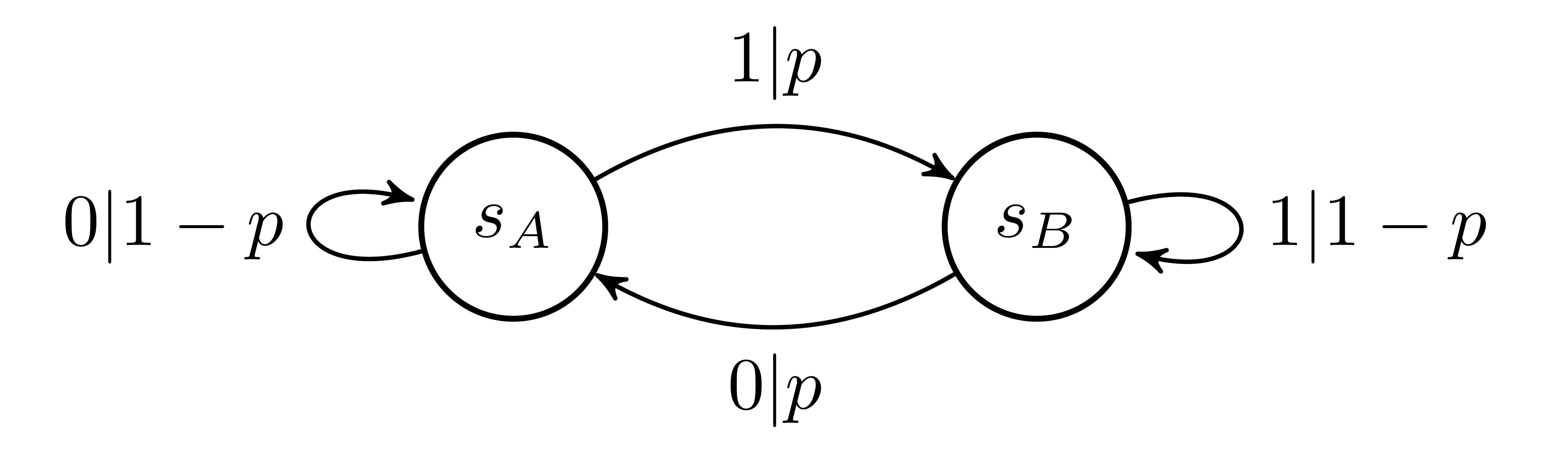}
	\caption{The perturbed coin process is illustrated with two causal states $S_A$ and $S_B$ that output $x$ with some probability $p$ or $1-p$.}
	\label{fig.perturbedcoin_suppmat}
\end{figure}

At every time step, the stochastic process goes from $\overleftarrow{x}$ to $\overleftarrow{x}x $ to $\overleftarrow{x}xx$ and so on. Each $\overleftarrow{x}$ is assigned to an equivalence class $s_j$ while $\overleftarrow{x}x$ is assigned to equivalence class $s_{\lambda(x,j)}$ labelled with deterministic update function $\lambda(x,j)$. This implies that $s_j x$ leads to $s_\lambda(x,j)$. One can make the connection that there are transitions between equivalence classes and each transition produces some outputs $x$ with some probability $p$. Therefore, the stochastic process can be represented with an edge-emitting hidden Markov model with states represented by equivalence classes and emissions given by $x$. As an example, the perturbed coin process can be represented by the following causal model in Fig.~\ref{fig.perturbedcoin_suppmat}.

These edge-emitting hidden Markov models are known as $\varepsilon$-machines. Running the $\varepsilon$-machine to generate a statistically identical stochastic process requires some memory that is defined by the Shannon entropy over the stationary distribution of causal states, also known as the statistical complexity of the stochastic process,
\begin{equation}
	C_\mu := H[P(s_j)] = -\sum_j P(s_j)\log_2P(s_j).
\end{equation}

\section{Tolerance for merging quantum memory states}
\label{suppmat.fidelityQMS}

The quantum inference protocol takes in a finite-length stochastic process of length $N$ and parses through the process with with $L$-length moving window, simulating the observation of $L$-length histories $x_{-L:0}$, to construct the following inferred quantum memory, $\ket{\varsigma_{x_{-L:0}}}$,
\begin{equation}
\label{eq.inferredQMS_suppmat}
	U \ket{\varsigma_{x_{-L:0}}} \ket{0} := \sum_{x_0\in\mathcal{A}} \sqrt{\tilde{P}(x_0|x_{-L:0})} \ket{\varsigma_{x_{-L+1:1}}} \ket{x_0}.
\end{equation}
The unitary operation acts on the memory state $\ket{\varsigma_{x_{-L:0}}}$ with a blank ancillary qubit $\ket{0}$ and transforms the memory state to $\ket{\varsigma_{x_{-L+1:1}}}$ while outputting $\ket{x}$ (see also Fig.~\ref{fig.decomposition}(a)). Repeated unitary operations with blank ancillary qubits produce an output string that should be statistically faithful to the input stochastic process with the caveat that $L$ is at least the Markov order and the length of the stochastic process is long enough.

The cardinality of output alphabets, $|\mathcal{A}|$, indicates that the quantum inference protocol produces a maximum of $|\mathcal{A}|^L$ quantum memory states for $L$-length history sequence of observables of the past. The corresponding size of the unitary operator has to be $|\mathcal{A}|^{L+1}$ by $|\mathcal{A}|^{L+1}$ to account for the appended ancillary qubit $\ket{0}$. Theoretically, constructing the unitary operator of size $|\mathcal{A}|^{L+1}$ by $|\mathcal{A}|^{L+1}$ is possible but it would be practically infeasible when applying on a quantum computer.
A large matrix tends to have large circuit depth after decomposing into implementable elementary quantum gates. It is imperative to minimise the dimension of the unitary operator for practicality purposes. Therefore, we propose merging statistically similar quantum memory states to reduce the state space, thereby ensuring that the number of qubits required is within acceptable and implementable means.

We compute the fidelity between two quantum memory states $\ket{\varsigma_{x_{-L:0}}}$ and $\ket{\varsigma_{x'_{-L:0}}}$ by taking the absolute value of their inner products, $|\braket{\varsigma_{x_{-L:0}}|\varsigma_{x'_{-L:0}}}|$. If the two quantum memory states are identical, then $|\braket{\varsigma_{x_{-L:0}}|\varsigma_{x'_{-L:0}}}| = 1$. We want to merge two identical quantum memory states and assign the same label, i.e. if $|\braket{\varsigma_{x_{-L:0}}|\varsigma_{x'_{-L:0}}}| = 1$, then $\ket{\varsigma_{x_{-L:0}}}, \ket{\varsigma_{x'_{-L:0}}} \mapsto \ket{\varsigma_i}$. However, the probability amplitudes in Eq.~\eqref{eq.inferredQMS_suppmat} are not exact probability amplitudes due to statistical fluctuations from having a finite stochastic process of length $N$. It is noted that these probability amplitudes will tend towards the exact values as $N_\text{process} \to \infty$.

As such, we define an equivalence relation in terms of the fidelity between two quantum memory states taking into account the $\delta$ factor as follows,
\begin{equation}
	\ket{\varsigma_{x_{-L:0}}} \sim \ket{\varsigma_{x'_{-L:0}}} \iff |\braket{\varsigma_{x_{-L:0}}|\varsigma_{x'_{-L:0}}}| \geq 1-\delta.
\end{equation}

The factor $\delta$ should be a function of $N_\text{process}$; small stochastic processes will incur greater errors to the inferred conditional probabilities. As such, we solve for the inner products for two quantum memory states $\braket{\varsigma_{x_{-L:0}}|\varsigma_{x'_{-L:0}}}$. Since the probability amplitudes in Eq. \eqref{eq.inferredQMS_suppmat} are real and positive, the absolute signs can be dropped.

\begin{equation}
\begin{aligned}
\label{eq.innerproducts_suppmat}
	\braket{\varsigma_{x_{-L:0}}|\varsigma_{x'_{-L:0}}} &= \bra{\varsigma_{x_{-L:0}}} U^\dagger U \ket{\varsigma_{x_{-L:0}'}} \\
	&= \sum_{x_0 \in \mathcal{A}} \sqrt{ \tilde{P}(x_{0}|x_{-L:0}) \tilde{P}(x_{0}|x_{-L:0}') } \braket{\varsigma_{x_{-L+1:1}}|\varsigma_{x'_{-L+1:1}}}.
\end{aligned}
\end{equation}

Iteratively applying Eq. \eqref{eq.innerproducts_suppmat}, we obtain
\begin{equation}
\begin{aligned}
	\braket{\varsigma_{x_{-L:0}}|\varsigma_{x'_{-L:0}}} &=  \sum_{x_{0:L}} \sqrt{ \tilde{P}(x_{0:L}|x_{-L:0}) \tilde{P}(x_{0:L}|x_{-L:0}') }.
\end{aligned}
\end{equation}

Inferred probability distributions are typically imperfect distributions with these imperfections arising from a finite data set. These imperfect probability distributions can be seen as a perturbed probability distribution. 

Suppose $\tilde{P} = P + \Delta P$ where $P$ is the exact distribution, and $\Delta P$ is some perturbation to the exact distribution, the square-root of two perturbed distributions $\tilde{P}$ and $\tilde{Q}$ can be simplified as,
\begin{equation}
\begin{aligned}
	\sqrt{\tilde{P} \tilde{Q}} &= \sqrt{ (P + \Delta P) (Q + \Delta Q) } \\
	&= \sqrt{ PQ + P \Delta Q + Q \Delta P + \Delta P \Delta Q} \\
	&= \sqrt{PQ} \left[ 1 + \frac{1}{2} \left( \frac{\Delta P}{P} + \frac{\Delta Q}{Q} + \frac{\Delta P \Delta Q}{PQ} \right) + ... \right] \\
	&\approx \sqrt{PQ} + \frac{\sqrt{PQ}}{2} \left[ \frac{\Delta P}{P} + \frac{\Delta Q}{Q} \right].
\end{aligned}
\end{equation}

As such, 
\begin{equation}
\begin{aligned}
	\braket{\varsigma_{x_{-L:0}}|\varsigma_{x'_{-L:0}}} &\approx \sum_{x_{0:L}} \sqrt{P(x_{0:L}|x_{-L:0}) P(x_{0:L}|x'_{-L:0})} \\ 
	& \hspace{2em} + \frac{\sqrt{P(x_{0:L}|x_{-L:0}) P(x_{0:L}|x'_{-L:0})}}{2} \left[ \frac{\Delta P(x_{0:L}|x_{-L:0})}{P(x_{0:L}|x_{-L:0})} + \frac{\Delta P(x_{0:L}|x'_{-L:0})}{P(x_{0:L}|x'_{-L:0})} \right].
\end{aligned}
\end{equation}

A randomly selected $L$-length word $x_{-L:0}$ is either $x_{0:L}$ or not, hence $P(x_{0:L}|x_{-L:0})$ follows a Bernoulli distribution. It is approximated to a normal distribution when the stochastic process is of large length $N$. As such, the standard deviation $\sigma_{P(x_{0:L}|x_{-L:0})} = \sqrt{P(x_{0:L}|x_{-L:0}) (1-P(x_{0:L}|x_{-L:0}))}$. We use the standard error of the mean to estimate the error for the mean value of $P(x_{0:L}|x_{-L:0})$ as follows $\sigma_{\bar{P}(x_{0:L}|x_{-L:0})} = \sqrt{ \frac{P(x_{0:L}|x_{-L:0}) (1-P(x_{0:L}|x_{-L:0}))}{N_\text{process}} }$.

Now, the magnitude of error $|\Delta P(x_{0:L}|x_{-L:0})|$ can be approximated as $|\sigma_{\bar{P}(x_{0:L}|x_{-L:0})}|$ to relate the existence of $\Delta P(x_{0:L}|x_{-L:0})$ to finite $N_\text{process}$,
\begin{equation}
\begin{aligned}
	|\Delta P(x_{0:L}|x_{-L:0})| &\approx |\sigma_{\bar{P}(x_{0:L}|x_{-L:0})}| \\
	&\approx \sqrt{ \frac{P(x_{0:L}|x_{-L:0}) (1-P(x_{0:L}|x_{-L:0}))}{N_\text{process}} }.
\end{aligned}
\end{equation}

The magnitude of the error to the overlaps between quantum memory states, which we denote as $\left|\Delta \braket{\varsigma_{x_{-L:0}}|\varsigma_{x'_{-L:0}}}\right|$, is
\begin{equation}
\begin{aligned}
	\left|\Delta \braket{\varsigma_{x_{-L:0}}|\varsigma_{x'_{-L:0}}}\right| &= \sum_{x_{0:L}} \frac{1}{2 \sqrt{N_\text{process}}} \left[ \frac{\left|\sqrt{P(x_{0:L}|x_{-L:0})(1-P(x_{0:L}|x_{-L:0}))}\right|}{\sqrt{P(x_{0:L}|x_{-L:0})}} + \frac{\left|\sqrt{P(x_{0:L}|x'_{-L:0})(1-P(x_{0:L}|x'_{-L:0}))}\right|}{\sqrt{P(x_{0:L}|x_{-L:0})}} \right] \\
\end{aligned}
\end{equation}

For $L$-length past and future, the conditional probabilities $\sum_{x_{0:L}} P(x_{0:L}|x_{-L:0})$ scale as $|\mathcal{A}|^L$ by $1$. Assuming that the conditional probabilities are almost equiprobable, then we can approximate $P(x_{0:L}|x_{-L:0}) \approx \frac{1}{|\mathcal{A}|^L}$. Keeping $L$ fixed, it is evident that $\left|\Delta \braket{\varsigma_{x_{-L:0}}|\varsigma_{x'_{-L:0}}}\right|$ scales proportionally to $\frac{1}{\sqrt{N_\text{process}}}$, i.e. the error to the inner products $\left|\Delta \braket{\varsigma_{x_{-L:0}}|\varsigma_{x'_{-L:0}}}\right|$ is directly related to $N_\text{process}$, the length of the stochastic process. As such, in our work, we set the dependence of $\delta$-tolerance directly on the length of the stochastic process $N_\text{process}$, to give $\delta = \frac{1}{2\sqrt{N_\text{process}}}$ with the factor of 2 giving a more stringent tolerance for merging quantum memory states.

\section{Generating unitaries}
\label{suppmat.generatingunitaries}

The procedure for constructing an appropriate unitary operator for a set of quantum memory sets comprises of four main steps~\cite{Binder2018}:
\begin{itemize}
	\item[(1)] Compute the inner products $c_{ij} = \braket{\sigma_i|\sigma_j}$.
	\item[(2)] Express the set of quantum memory states $\{\ket{\sigma_j}\}$ in a chosen orthogonal basis $\{\ket{e_j}\}$ in Hilbert space of dimensions $|\sigma|$ using the inverse Gram-Schmidt procedure that gives the following set of equations
	\begin{equation}
	\begin{aligned}
		\ket{\sigma_1} &= \ket{e_1} \\
		\ket{\sigma_2} &= c_{12}\ket{e_1} + \sqrt{1-c_{12}^2} \ket{e_2} \text{, etc.}
	\end{aligned}
	\end{equation}
	\item[(3)] Solve for matrix elements $U_{ix,i'0} = \bra{e_i} \bra{x} U \ket{e_{i'}} \ket{x=0}$. The $\ket{x=0}$ factor represents the blank ancillary state and it fills in the odd-numbered columns.
	\item[(4)] Fill in the remaining even-numbered columns of $U$ with a Gram-Schmidt procedure with the already-determined columns in Step (3). For example, one can assign random numbers to the even-numbered columns then proceed with the Gram-Schmidt procedure.
\end{itemize}

The same four steps can be used to construct the unitary operator for the set of merged inferred quantum states obtained with the quantum inference protocol, $\{\ket{\varsigma_i}\}$. 

The columns of the unitary operator with $x=0$ (or odd numbered columns) are uniquely defined as follows,
\begin{equation}
\label{eq.UnitaryElement1}
	U_{ix,i'0} \equiv \bra{e_i}\bra{x}U\ket{e_{i'}}\ket{x=0}
\end{equation}
with the set of $\ket{e_i}$ being the basis states obtained from a forward Gram-Schmidt procedure with the merged and relabelled set of quantum memory states $\{\ket{\varsigma_i}\}$. Any basis state $\ket{e_i}$ can be rewritten as a linear combination of the merged quantum memory states. For example,
\begin{equation}
\begin{aligned}
\label{eq.linearcombination}
	\ket{e_{i=1}} &= \gamma_{i=1,k=1} \ket{\varsigma_{k=1}} \\
	\ket{e_{i=2}} &= \gamma_{i=2,k=1} \ket{\varsigma_{k=1}} + \gamma_{i=2,k=2} \ket{\varsigma_{k=2}} \\
	\ket{e_{i=3}} &= \gamma_{i=3,k=1} \ket{\varsigma_{k=1}} + \gamma_{i=3,k=2} \ket{\varsigma_{k=2}} + \gamma_{i=3,k=3} \ket{\varsigma_{k=3}}
\end{aligned}
\end{equation}
with $\gamma_{i,k}$ being the respective coefficients that can be found with the Gram-Schmidt procedure. Substituting Eq.~\eqref{eq.linearcombination} into Eq.~\eqref{eq.UnitaryElement1}, each element within the unique columns of the unitary operator is given by
\begin{equation}
\label{eq.UnitaryElement2}
	U_{ix,i'0} = \sum_{k=1}^{k=i} \sum_{k'=1}^{k'=i'} (\gamma_{i,k}) (\gamma_{i',k'}) \sqrt{P(x|\varsigma_{k'})} \braket{\varsigma_i|\varsigma_{\lambda(x,\varsigma_{k'})}}.
\end{equation}
After these columns have been filled, the remaining columns can be found by assigning all entries with some random numbers then orthonormalised with respect to the odd numbered columns that were obtained earlier through a Gram-Schmidt procedure.

The resultant quantum memory states for which the unitary acts on dictates that the overlaps $\braket{\varsigma_i|\varsigma_j}$ be preserved. We can thus rewrite the first quantum memory state as $\ket{\varsigma_1} = [1,0,0...]^T$, the second as $\ket{\varsigma_2} = [\braket{\varsigma_1|\varsigma_2},\sqrt{1-\braket{\varsigma_1|\varsigma_2}^2},0...]^T$, and so on
\footnote{The unitary reconstruction algorithm has been made available on GitHub at \url{https://github.com/matthew0021/unitary-construction-and-decomposition}.}.

\section{Decomposing unitaries}
\label{suppmat.decomposingunitaries}

It is vital that any unitary operator is first decomposed into elementary quantum gates so that it can be applied on a quantum computer. There exist several decomposition methods \cite{VanLoan1985, Bai1993, Paige1994, Mottonen2004, Mottonen2005, Chen2013, Nakajima2006, Fuhr2018}, some of which have been automated with codes. Although Qiskit -- the platform we use for numerical work -- contains its own version of 2 qubit unitary decomposition~\footnote{Two Qubit Basis Decomposer. \url{https://qiskit.org/documentation/stubs/qiskit.quantum_info.TwoQubitBasisDecomposer.html}. Last Accessed: 2021-03-30}, we opt not to use it for two reasons. It is limited to two qubits but more importantly, our operator is not guaranteed to be a numerically exact unitary operator where $U^\dagger U = I$ after taking statistical fluctuations into account.

We will use the cosine-sine decomposition~\cite{Paige1994} coupled with generalised singular value decomposition~\cite{VanLoan1985} for our decomposition algorithm.

The general cosine-sine decomposition states that any $2^n$ by $2^n$ matrix can be split into four block matrices $g_{11}, g_{12}, g_{21}, g_{22}$ of size $2^{n-1}$ by $2^{n-1}$. Singular value decomposition is then applied to any two of these four block matrices to obtain the following decomposition,	
\begin{equation}
\begin{aligned}
\label{eq.gsvdcsd}
	U &\equiv \begin{bmatrix} g_{11} & g_{12} \\ g_{21} & g_{22} \end{bmatrix} = \begin{bmatrix} u_1 &   \\   & u_2 \end{bmatrix} \begin{bmatrix} c & -s \\ s & c \end{bmatrix} \begin{bmatrix} v_1 &   \\   & v_2 \end{bmatrix}
\end{aligned}		
\end{equation}
where
\begin{equation}
\begin{aligned}
	g_{11} &= u_1 c v_1;   \hspace{0.6cm} &&g_{12} = - u_1 s v_2; \\
	g_{21} &= u_2 s v_1; \hspace{0.6cm} &&g_{22} = u_2 c v_2.
\end{aligned}
\end{equation}
Spaces in matrices are square matrices consisting of all zero entries of the appropriate dimensions.

The middle term in Eq.~\eqref{eq.gsvdcsd} known as a multiplexor gate~\cite{Mottonen2004} and constitutes a matrix with diagonal cosine and sine terms,
\begin{equation}
\begin{aligned}
	c &\equiv \text{diag}(\cos\theta_1, \cos\theta_2,...,\cos\theta_{2^{n-1}}) \\
	s &\equiv \text{diag}(\sin\theta_1, \sin\theta_2,...,\sin\theta_{2^{n-1}}).
\end{aligned}
\end{equation}
A multiplexor gate is one which allows `if-else' conditions. For example, a CNOT gate is a prime example of a multiplexor gate. If the control qubit is $\ket{0}$, the state of the target qubit is not changed. Else (if the control qubit is $\ket{1}$), the state of the target qubit is flipped.

For two qubits multiplexors with arbitrary angles, $4$ by $4$ multiplexor gate contains two angles $\theta_1$ and $\theta_2$ where a Y-rotation with angle $\theta_1$ is applied to the first qubit if the second qubit is $\ket{0}$ and a Y-rotation with angle $\theta_2$ is applied to the first qubit if the second qubit is $\ket{1}$, i.e.
\begin{equation}
	\begin{bmatrix} c & -s \\ s & c \end{bmatrix} \ket{q_0} \ket{q_1} = \left\{
		\begin{array}{ll}
		(R_y(\theta_1) \otimes id) \ket{q_0} \ket{q_1} \text{, if } \ket{q_1} = \ket{0} \\
		(R_y(\theta_2) \otimes id) \ket{q_0} \ket{q_1} \text{, if } \ket{q_1} = \ket{1} 
		\end{array}
	\right.
\end{equation}
To further simplify the left and right matrices in Eq.~\eqref{eq.gsvdcsd}, we use the matrix identity \cite{DeVos2016}
\begin{equation}
\label{eq.matrixidentity1}
	\begin{bmatrix} \alpha &   \\   & \beta \end{bmatrix} = \begin{bmatrix}   & I \\ I &   \end{bmatrix} \begin{bmatrix} I &   \\   & \alpha \end{bmatrix} \begin{bmatrix}   & I \\ I &   \end{bmatrix} \begin{bmatrix} I &   \\   & \beta \end{bmatrix},
\end{equation}
with $\alpha, \beta \in U(2^{n-1})$ and $I$ is a $2^{n-1}$ by $2^{n-1}$ identity matrix. Furthermore,
\begin{equation}
	\begin{bmatrix}   & I \\ I &   \end{bmatrix} \equiv X \otimes I 
\end{equation}
with $X$ being the NOT gate.

The left and right matrices of Eq.~\eqref{eq.gsvdcsd} can be simplified by repeatedly applying singular value decomposition to $u_1, u_2, v_1, v_2$ until a 2 by 2 unitary operator $U(2)$ is achieved. Once the single qubit $U(2)$ matrices are obtained, they can be decomposed into either ZYZ (or ZXZ) rotations with some angles $\gamma, \phi, \theta, \lambda$ \cite{Nielsen2010},
\begin{equation}
\label{eq.ZYZdecomposition}
	U = e^{i\gamma} R_Z(\phi) R_Y(\theta) R_Z(\lambda).
\end{equation}

An arbitrary $2^n$ by $2^n$ unitary operator which is decomposed with the cosine-sine decomposition and applied for $t$-times in a quantum circuit has circuit depth 
\begin{equation}
    t\left[7(4^{n-1}) + 5\left(\sum_{i=0}^{n-2} 4^i\right)\right].
\end{equation}

\section{Quantum Error Mitigation}
\label{suppmat.quantumerrormitigation}

\textbf{Effect of finite $N_\text{GST}$-shots for gate set tomography.} Finite $N_\text{GST}$ shots for gate set tomography affects a few variables that contribute to the overall accuracy of the error mitigation. These variables will be written as a sum of the exact value and a fixed perturbation term $\Delta$ with scaling parameter $\epsilon$ to denote the strength of the perturbation.
\begin{itemize}
	\item[(1)] Gram matrix $g$: Let $g_{N_\text{GST}}$ be the Gram matrix with $N_\text{GST}$ shots. Thus, $g_{N_\text{GST}} = g + \epsilon \Delta g$.
	\item[(2)] $\tilde{\mathcal{O}}^{(l)}$: Let $\tilde{\mathcal{O}}^{(l)}_{N_\text{GST}} = \tilde{\mathcal{O}}^{(l)} + \epsilon \Delta \tilde{\mathcal{O}}^{(l)}$.
	\item[(3)] Basis operations $\tilde{B}_i$. Let $\tilde{B}_{i_{N_\text{GST}}} = \tilde{B}_i + \epsilon \Delta \tilde{B}_i^{(l)}$.
\end{itemize}
Our goal is to provide intuition of how the effects of finite $N_\text{GST}$ shots in the pre-experimental portion cascade down to the cost of error mitigation. We begin with
\begin{equation} 
	\hat{\mathcal{O}}^{(l)} = T g^{-1} \tilde{\mathcal{O}}^{(l)} T^{-1}.
\end{equation}
For $\hat{\mathcal{O}}^{(l)}$ being affected by $N_\text{GST}$, we have
\begin{equation}
\begin{aligned}
\label{eq.perturbedOhat_suppmat1}
	\hat{\mathcal{O}}^{(l)}_{N_\text{GST}} &= T g_{N_\text{GST}}^{-1} \tilde{\mathcal{O}}^{(l)}_{N_\text{GST}} T^{-1} \\
	&= T \left( g + \epsilon \Delta g \right)^{-1} \left( \tilde{\mathcal{O}}^{(l)} + \epsilon \Delta \tilde{\mathcal{O}}^{(l)} \right) T^{-1} \\
	&= T \left( g^{-1} - g^{-1} \epsilon \Delta g g^{-1} \right) \left( \tilde{\mathcal{O}}^{(l)} + \epsilon \Delta \tilde{\mathcal{O}}^{(l)} \right) T^{-1} \\
	&= T g^{-1} \tilde{\mathcal{O}}^{(l)} T^{-1} + T g^{-1} \epsilon \Delta \tilde{\mathcal{O}}^{(l)} T^{-1} - T g^{-1} \epsilon \Delta g g^{-1} \tilde{\mathcal{O}}^{(l)} T^{-1} + O(\epsilon^2). \\
\end{aligned}
\end{equation}
This gives
\begin{equation}
\begin{aligned}
\label{eq.perturbedOhat_suppmat2}
	\hat{\mathcal{O}}^{(l)} &= T g^{-1} \tilde{\mathcal{O}}^{(l)} T^{-1} \\
	\epsilon \Delta \hat{\mathcal{O}}^{(l)} &= \epsilon \left[ T g^{-1} \Delta \tilde{\mathcal{O}}^{(l)} T^{-1} - T g^{-1} \Delta g g^{-1} \tilde{\mathcal{O}}^{(l)} T^{-1} \right] + O(\epsilon^2).
\end{aligned}
\end{equation}
From this, we get $\|\epsilon\Delta \hat{\mathcal{O}}^{(l)}\|\sim \epsilon \left(\|\Delta \tilde{\mathcal{O}}^{(l)}\|+\|\Delta g\|\right)$ up to the first order of $\epsilon$, where $\|\cdot\|$ is the operator norm. The error to $\epsilon \Delta \hat{\mathcal{O}}^{(l)}$ has contributions from $\epsilon \Delta \tilde{\mathcal{O}}^{(l)}$ and $\epsilon \Delta g$. More concretely, since the strength of perturbation is largely dominated by how many $N_\text{GST}$ shots is performed, we set $\epsilon = 1$ and let $\Delta$ represent the full extent perturbation due to $N_\text{GST}$ shots. The error for basis operations $\hat{\mathcal{B}}_i$ can similarly be found by replacing the operators $\tilde{\mathcal{O}}^{(l)}$ in Eq.~\eqref{eq.perturbedOhat_suppmat1}.

The inverse of noise can be found as
\begin{equation}
	\left( \mathcal{N}^{(l)} \right)^{-1} = \mathcal{O}^{(l),\text{exact}} \left( \hat{\mathcal{O}}^{(l)} \right)^{-1}
\end{equation}
which means that for $N_\text{GST}$, the inverse noise $\left( \mathcal{N}_{N_\text{GST}}^{(l)} \right)^{-1}$ is
\begin{equation}
\begin{aligned}
\label{eq.invNoise_GST_suppmat}
	\left( \mathcal{N}_{N_\text{GST}}^{(l)} \right)^{-1} &= \mathcal{O}^{(l),\text{exact}} \left( \hat{\mathcal{O}}_{N_\text{GST}}^{(l)} \right)^{-1} \\
	&= \mathcal{O}^{(l),\text{exact}} \left( \left( \hat{\mathcal{O}}^{(l)} \right) + \Delta  \left( \hat{\mathcal{O}}^{(l)} \right) \right)^{-1} \\
	&\approx \mathcal{O}^{(l),\text{exact}} \left( \left( \hat{\mathcal{O}}^{(l)} \right)^{-1} - \left( \hat{\mathcal{O}}^{(l)} \right)^{-1} \Delta  \hat{\mathcal{O}}^{(l)} \left( \hat{\mathcal{O}}^{(l)} \right)^{-1} \right) \\
	&= \mathcal{O}^{(l),\text{exact}} \left( \hat{\mathcal{O}}^{(l)} \right)^{-1} - \mathcal{O}^{(l),\text{exact}} \left( \hat{\mathcal{O}}^{(l)} \right)^{-1} \Delta  \hat{\mathcal{O}}^{(l)} \left( \hat{\mathcal{O}}{(l)} \right)^{-1}
\end{aligned}
\end{equation}
This implies $\|\Delta(\mathcal{N}^{(l)})^{-1}\|\sim \|\Delta \hat{\mathcal{O}}^{(l)}\|$, and together with \eqref{eq.perturbedOhat_suppmat2}, we get $\|\Delta(\mathcal{N}^{(l)})^{-1}\|\sim\|\Delta \tilde{\mathcal{O}}^{(l)}\|+\|\Delta g\|$.

Should one print out the values of $\left( \mathcal{N}^{(l)} \right)^{-1}$ for different values of $N_\text{GST}$, one will notice that the elements of $\left( \mathcal{N}^{(l)} \right)^{-1}$ will increase with smaller $N_\text{GST}$, especially for the off-diagonal elements.  This shows the effects of having finite $N_\text{GST}$ shots.

We expect the quasiprobabilities in the quasiprobability decomposition from $N_\text{GST}$ to fluctuate as well. Hence, we let each quasiprobability with index $i$ to be $q_{\mathcal{O}^{(l)},{N_\text{GST}},i} = q_{\mathcal{O}^{(l)},i} + \Delta q_{\mathcal{O}^{(l)},i}$. To first order of $\Delta$,
\begin{equation}
\begin{aligned}
	\left( \mathcal{N}^{(l)}_\text{GST} \right)^{-1} &= \sum_i q_{\mathcal{O}^{(l)},{N_\text{GST}},i} \hat{\mathcal{B}}_{N_\text{GST},i} \\
	\left( \mathcal{N}^{(l)} + \Delta \mathcal{N}^{(l)} \right)^{-1} &= \sum_i \left( q_{\mathcal{O}^{(l)},i} + \Delta q_{\mathcal{O}^{(l)},i} \right) \left( \hat{\mathcal{B}}_i + \Delta \hat{\mathcal{B}}_i \right) \\
	\left( \left( \mathcal{N}^{(l)} \right)^{-1} - \left( \mathcal{N}^{(l)} \right)^{-1} \Delta \mathcal{N}^{(l)} \left( \mathcal{N}^{(l)} \right)^{-1} \right) &\approx \sum_i q_{\mathcal{O}^{(l)},i} \hat{\mathcal{B}}_i + q_{\mathcal{O}^{(l)},i} \Delta \hat{\mathcal{B}}_i + \Delta q_{\mathcal{O}^{(l)},i} \hat{\mathcal{B}}_i .
\end{aligned}
\end{equation}
The error from $\Delta \mathcal{N}^{(l)}$ contributes to $q_{\mathcal{O}^{(l)},i}$ by a factor of $\Delta \hat{\mathcal{B}}_i$ but more importantly incurs a perturbation $\Delta q_{\mathcal{O}^{(l)},i}$. The fluctuation to $C_{\mathcal{O}^{(l)}}$ is hence $\Delta C_{\mathcal{O}^{(l)}} = \sum_i |\Delta q_{\mathcal{O}^{(l)},i}|$.

\vspace{1cm}

\textbf{Scaling with respect to $N_\text{GST}$.} We are now in a position to ascertain each variable scales with respect to $N_\text{GST}$.  The matrix $g$ has elements $g_{j,k}$ that are comprised of measuring an initialised state $\rho_k$ in basis $M_j$ which yields outcomes $0$ or $1$, allowing it to be represented by a Bernoulli distribution. The argument holds for $\tilde{\mathcal{O}}^{(l)}$ (and $\tilde{\mathcal{B}}_i$) as well, where its matrix has elements $\tilde{\mathcal{O}}^{(l)}_{j,k}$ are built from outcomes $0$ or $1$ with some probabilities $p$ or $1-p$. Each matrix element is approximated to a normal distribution when $N_\text{GST}$ is large enough, allowing the standard deviation to be $\sigma_p = \sqrt{p(1-p)}$. Invoking the standard error of the mean is used to estimate the error for the expectation value, $\sigma_{\bar{p}} = \sqrt{ \frac{p(1-p)}{N_\text{GST}} }$, $\|\Delta g\|$, $\|\Delta \tilde{\mathcal{O}}^{(l)}\|$, and $\|\Delta \tilde{\mathcal{B}}_i\|$ all scale as $1/\sqrt{N_\text{GST}}$. This implies that $\|\Delta \left( \hat{\mathcal{B}}_i \right)^{-1}\| \approx \|\Delta \left( \mathcal{N}^{(l)} \right)^{-1}\| \approx 2/\sqrt{N_\text{GST}}$. And hence, $\Delta q_{\mathcal{O}^{(l)},i}$ scales roughly as $2/\sqrt{N_\text{GST}}$. The eventual error to $C_{\mathcal{O}^{(l)}}$ is $\Delta C_{\mathcal{O}^{(l)}} = \sum_i | \Delta q_{\mathcal{O}^{(l)},i} |$.

\vspace{1cm}

\textbf{Post-processing for multi-step error mitigation.}
We showed in the main text that post-processing the Monte Carlo results yields the following estimated error mitigated expectation values,
\begin{equation}
\begin{aligned}
	P_\text{QEM}(x=0) &= C \left( \frac{\text{No. of }x = 0 | \text{sgn} = +1}{N_\text{MC}} - \frac{\text{No. of }x = 0 | \text{sgn} = -1}{N_\text{MC}} \right) \\
	P_\text{QEM}(x=1) &= C \left( \frac{\text{No. of }x = 1 | \text{sgn} = +1}{N_\text{MC}} - \frac{\text{No. of }x = 1 | \text{sgn} = -1}{N_\text{MC}} \right)
\end{aligned}
\end{equation}
where
\begin{equation}
\begin{aligned}
	\text{sgn} &= \text{sign} \left( q_{\rho,k} [ \prod_l q_{\mathcal{O}^{(l)}} ] q_{M,j} \right) \\
	&= \text{sign} \left( q_{\rho,k} [ q_{\mathcal{O}^{(1)}} q_{\mathcal{O}^{(2)}} ... q_{\mathcal{O}^{(l)}} ] q_{M,j} \right).
\end{aligned}
\end{equation}
and
\begin{equation}
\begin{aligned}
\label{eq.C_coefficient}
	C &= C_{\rho} [ \prod_l C_{\mathcal{O}^{(l)}} ] C_M \\
	&= C_{\rho} [ C_{\mathcal{O}^{(1)}} C_{\mathcal{O}^{(2)}} ... C_{\mathcal{O}^{(l)}} ] C_M. 
\end{aligned}
\end{equation}

The same methodology can be extended to multi-steps. For instance, the 2-step estimated error mitigated expectation values are given by
\begin{equation}
\begin{aligned}
    P_\text{QEM}(x^{(t=1)}=0, x^{(t=2)}=0) &= C \left[ \frac{\text{No. of } x^{(t=1)} x^{(t=2)} = 00 | \text{sgn} = +1}{N} - \frac{\text{No. of } x^{(t=1)} x^{(t=2)} = 00 | \text{sgn} = -1}{N} \right] \\
    P_\text{QEM}(x^{(t=1)}=0, x^{(t=2)}=1) &= C \left[ \frac{\text{No. of } x^{(t=1)} x^{(t=2)} = 01 | \text{sgn} = +1}{N} - \frac{\text{No. of } x^{(t=1)} x^{(t=2)} = 01 | \text{sgn} = -1}{N} \right] \\
    P_\text{QEM}(x^{(t=1)}=1, x^{(t=2)}=0) &= C \left[ \frac{\text{No. of } x^{(t=1)} x^{(t=2)} = 10 | \text{sgn} = +1}{N} - \frac{\text{No. of } x^{(t=1)} x^{(t=2)} = 10 | \text{sgn} = -1}{N} \right] \\
    P_\text{QEM}(x^{(t=1)}=1, x^{(t=2)}=1) &= C \left[ \frac{\text{No. of } x^{(t=1)} x^{(t=2)} = 11 | \text{sgn} = +1}{N} - \frac{\text{No. of } x^{(t=1)} x^{(t=2)} = 11 | \text{sgn} = -1}{N} \right] .
\end{aligned}
\end{equation} 

We include superscripts $(t)$ to indicate the index of the multi-step distribution. One can easily extrapolate to find further multi-step error mitigated distributions.

\vspace{1cm}

\textbf{Basis operations.}
Deterministic channels acting on a single qubit require 13 basis operations~\cite{Takagi2020}. These basis operations are CPTP maps and are listed in Table \ref{table.1}. Deterministic channels acting on two qubits require $13^2 + 72 = 241$ basis operations. While $13^2$ can be straightforwardly found by taking the tensor product between each $\bar{\mathcal{B}}_i$ in Table \ref{table.1}, the remaining $72$ operations are made up of CNOTs, controlled-phase, controlled-Hadamards, CNOTs with eigenstates of the Hadamard gate, SWAP, and iSWAP gates. These 72 basis operations are listed in Table \ref{table.2}.

\begin{table}[h!]
	\centering
	\begin{tabular}{| c | c |} 
	 \hline
	 $\bar{\mathcal{B}}_1$	& 	$id$  \\ 
	 $\bar{\mathcal{B}}_2$	& 	$\mathcal{X}$  \\ 
	 $\bar{\mathcal{B}}_3$	& 	$\mathcal{Y}$  \\ 
	 $\bar{\mathcal{B}}_4$	& 	$\mathcal{Z}$  \\ 
	 $\bar{\mathcal{B}}_5$	& 	$\mathcal{H}^\dagger \mathcal{S}^\dagger \mathcal{H}$   \\ 
	 $\bar{\mathcal{B}}_6$	& 	$\mathcal{S} \mathcal{H} \mathcal{S}^\dagger \mathcal{H}^\dagger \mathcal{S}^\dagger$  \\ 
	 $\bar{\mathcal{B}}_7$	& 	$\mathcal{S}^\dagger$  \\ 
	 $\bar{\mathcal{B}}_8$	& 	$\mathcal{S} \mathcal{H} \mathcal{S}^\dagger$  \\ 
	 $\bar{\mathcal{B}}_9$	& 	$\mathcal{H}$  \\ 
	 $\bar{\mathcal{B}}_{10}$	& 	$\mathcal{H}^\dagger \mathcal{S}^\dagger \mathcal{H} \mathcal{S} \mathcal{H}$  \\ 
	 $\bar{\mathcal{B}}_{11}$	& 	$\mathcal{P}_{\ket{+}}$  \\ 
	 $\bar{\mathcal{B}}_{12}$	& 	$\mathcal{P}_{\ket{y+}}$  \\ 
	 $\bar{\mathcal{B}}_{13}$	& 	$\mathcal{P}_{\ket{0}}$  \\ 
	 \hline
	\end{tabular}
	\caption{The 13 CPTP maps listed here are quantum channels that can be used for basis operations. $id$ represents the identity channel while $\mathcal{P}_{\ket{\cdot}}$ signifies a state preparation channel that prepares the state $\ket{\cdot}$.}
	\label{table.1}
\end{table}

\begin{table}[h!]

	\centering
	\begin{tabular}{| c | c |} 
	 \hline
	 $\bar{\mathcal{B}}_{170 \text{to} 178}$	& 	$\mathcal{C}\mathcal{X}$ + conjugation with $\mathcal{K}_{1,2}, \mathcal{K}^\dagger_{1,2}$  \\ 
	 $\bar{\mathcal{B}}_{179 \text{to} 187}$	& 	$\mathcal{X}_1 \circ \mathcal{C}\mathcal{X} \circ \mathcal{X}_1$ + conjugation with $\mathcal{K}_{1,2}, \mathcal{K}^\dagger_{1,2}$  \\ 
	 $\bar{\mathcal{B}}_{188 \text{to} 196}$	& 	$\mathcal{C}\mathcal{S}$ + conjugation with $\mathcal{K}_{1,2}, \mathcal{K}^\dagger_{1,2}$  \\ 
	 $\bar{\mathcal{B}}_{197 \text{to} 205}$	& 	$\mathcal{C}\mathcal{H}$ + conjugation with $\mathcal{K}_{1,2}, \mathcal{K}^\dagger_{1,2}$   \\ 
	 $\bar{\mathcal{B}}_{206 \text{to} 178}$	& 	$\mathcal{C}_\mathcal{H}\mathcal{X}$ + conjugation with $\mathcal{K}_{1,2}, \mathcal{K}^\dagger_{1,2}$  \\ 
	 $\bar{\mathcal{B}}_{215 \text{to} 223}$	& 	$\mathcal{C}\mathcal{X} \circ \mathcal{H}_1$ + conjugation with $\mathcal{K}_{1,2}, \mathcal{K}^\dagger_{1,2}$  \\ 
	 $\bar{\mathcal{B}}_{224 \text{to} 226}$	& 	$\mathcal{S}_{swap}$ + conjugation with $\mathcal{K}_{2}, \mathcal{K}^\dagger_{2}$  \\ 
	 $\bar{\mathcal{B}}_{227 \text{to} 232}$	& 	$i\mathcal{S}_{swap}$ + conjugation with $\mathcal{K}_{1,2}, \mathcal{K}^\dagger_{2}$  \\ 
	 $\bar{\mathcal{B}}_{233 \text{to} 241}$	& 	$\mathcal{S}_{swap} \circ \mathcal{H}_1$ + conjugation with $\mathcal{K}_{1,2}, \mathcal{K}^\dagger_{1,2}$  \\ 
	 \hline
	\end{tabular}
	\caption{A system of $n=2$ qubits requires 241 basis operations. The first 169 can be obtained by taking the tensor product of the 13 basis operations listed in Table \ref{table.1} while the remaining 72 basis operations are listed here. Here, we let $\mathcal{K} = \mathcal{S}\circ\mathcal{H}$.}
	\label{table.2}
\end{table}

As a follow up to Table \ref{table.2}, ``$\mathcal{U}$ + conjugation with $\mathcal{K}_{1,2},\mathcal{K}^\dagger_{1,2}$" means:
\begin{equation}
\begin{aligned}
	& \big( \mathcal{K}_1 & \otimes & \mathcal{K}_2 \big) & \circ & \mathcal{U} & \circ & \big( \mathcal{K}^\dagger_1 & \otimes & \mathcal{K}^\dagger_2 \big) \\
	& \big( \mathcal{K}_1 & \otimes & \mathcal{K}^\dagger_2 \big) & \circ & \mathcal{U} & \circ & \big( \mathcal{K}^\dagger_1 & \otimes & \mathcal{K}_2 \big) \\
	& \big( \mathcal{K}_1 & \otimes & id \big) & \circ & \mathcal{U} & \circ & \big( \mathcal{K}^\dagger_1 & \otimes & id \big) \\
	& \big( \mathcal{K}^\dagger_1 & \otimes & \mathcal{K}_2 \big) & \circ & \mathcal{U} & \circ & \big( \mathcal{K}_1 & \otimes & \mathcal{K}^\dagger_2 \big) \\
	& \big( \mathcal{K}^\dagger_1 & \otimes & \mathcal{K}^\dagger_2 \big) & \circ & \mathcal{U} & \circ & \big( \mathcal{K}_1 & \otimes & \mathcal{K}_2 \big) \\
	& \big( \mathcal{K}^\dagger_1 & \otimes & id \big) & \circ & \mathcal{U} & \circ & \big( \mathcal{K}_1 & \otimes & id \big) \\
	& \big( id & \otimes & \mathcal{K}_2 \big) & \circ & \mathcal{U} & \circ & \big( id & \otimes & \mathcal{K}^\dagger_2 \big) \\
	& \big( id & \otimes & \mathcal{K}^\dagger_2 \big) & \circ & \mathcal{U} & \circ & \big( id & \otimes & \mathcal{K}_2 \big) \\
	& \big( id & \otimes & id \big) & \circ & \mathcal{U} & \circ & \big( id & \otimes & id \big) \\
\end{aligned}
\end{equation}

\section{Scaling of fidelity with respect to $N_\text{MC}$}
\label{sec.deltascaling}

Suppose $N_\text{MC}$ is large enough such that $\sigma_\text{MC}$ is small enough that the distribution of the estimated expectation values does not cross into ill-defined regions (where probability is less than 0 or greater than 1) where we could form a valid probability distribution. We can choose any quantifier to quantify the distance between the error mitigated distribution and the exact, such as the fidelity. How would such distance measure scale with respect to $N_\text{MC}$? 

Let $P_\text{ex}$ be the exact distribution and $P_\text{sim}$ be the distribution obtained through simulation. $P_\text{sim}$ denotes the distribution of error mitigated expectation values of the simulation as well as the noisy simulation that has no error mitigation. The fidelity between two distributions $P_\text{ex}$ and $P_\text{sim}$ is
\begin{equation}
	F(P_\text{ex},P_\text{sim}) = \sum_i \sqrt{(p_{\text{ex,i}}) (p_{\text{sim},i})}.
\end{equation}

Suppose $P_\text{sim}$ can be seen as a distribution that has some perturbation to $P_\text{ex}$.  Then we have the following equation to second order of $\Delta P_\text{ex}$,
\begin{equation}
\begin{aligned}
\label{eq.fidelity_perturbation1}
	F(P_\text{ex},P_\text{ex}+\Delta P_\text{ex}) &= \sum_i \sqrt{(p_{\text{ex},i}) (p_{\text{ex},i} + \Delta p_{\text{ex},i})} \\
	&= \sum_i p_{\text{ex},i} \sqrt{1 + \frac{\Delta p_{\text{ex},i}}{p_{\text{ex},i}}} \\
	&\approx \sum_i p_{\text{ex},i} \left[ 1 + \frac{1}{2} \left( \frac{\Delta p_{\text{ex},i}}{p_{\text{ex},i}} \right) - \frac{1}{8} \left( \frac{\Delta p_{\text{ex},i}}{p_{\text{ex},i}} \right)^2 \right] \\
	&= \sum_i \left[ p_{\text{ex},i} + \frac{1}{2} (\Delta p_{\text{ex},i}) - \frac{1}{8} \frac{(\Delta p_{\text{ex},i} )^2}{p_{\text{ex},i}} \right].
\end{aligned}
\end{equation}
The conservation of probability dictates that $\sum_i p_{\text{ex},i} = p_{\text{sim},i} = 1$ it can be observed that $\sum_i \Delta p_{\text{ex},i} = 0$. Because fidelity is bounded by $0 \leq F \leq 1$~\cite{Jozsa1994Fidelity}, Therefore, Eq.~\eqref{eq.fidelity_perturbation1} resolves to
\begin{equation}
\begin{aligned}
	F(P_\text{ex},P_\text{ex}+\Delta P_\text{ex}) &= \sum_i \left[ p_{\text{ex},i} + \frac{1}{2} (\Delta p_{\text{ex},i}) - \frac{1}{8} \frac{(\Delta p_{\text{ex},i} )^2}{p_{\text{ex},i}} \right] \\
	&= 1 + 0 - \sum_i \frac{1}{8} \frac{(\Delta p_{\text{ex},i} )^2}{p_{\text{ex},i}} \\
&= F(P_\text{ex},P_\text{ex}) - \Delta F
\end{aligned}
\end{equation}
where $\Delta F = \sum_i \frac{1}{8} \frac{(\Delta p_{\text{ex},i} )^2}{p_{\text{ex},i}}$. For each time step $t$ of unitary operation, the error mitigation technique's expectation values for each $p_{\text{sim},i}$ occurs with a standard deviation $\sigma^{(t)}_\text{MC} = \frac{C^t}{\sqrt{N_\text{NC}}}$. As such, $\Delta p_i = \frac{C^t}{\sqrt{N_\text{MC}}}$. Substituting this, we obtain the scaling for $\Delta F$ as
\begin{equation}
\begin{aligned}
	\Delta F &= \sum_i \frac{1}{8 p_{\text{ex},i}} \left( \frac{C^t}{\sqrt{N_\text{MC}}} \right)^2 \\
	&= \frac{C^{2t}}{8 N_\text{MC}} \sum_i \frac{1}{p_{\text{ex},i}} .
\end{aligned}
\end{equation}

Therefore, the perturbation to the fidelity scales as $\sim \frac{C^{2t}}{N_\text{MC}}$.

\end{document}